\documentclass{aastex}
\usepackage[onecolumn]{emulateapj5}

\newcommand{\Msun}{\mbox{M$_\odot$}}		 
\newcommand{\Msunyr}{\mbox{M$_\odot$~yr$^{-1}$}}	 
\newcommand{\WFC}{\mbox{\sl WFC\/}}
\newcommand{\WFPCTwo}{\mbox{\sl WFPC2\/}}
\newcommand{\HST}{\mbox{\sl HST\/}}

\slugcomment{\today}
\shorttitle{HST/WFPC2 Study of the Trapezium Cluster}
\shortauthors{Robberto et al.}
\received{13 December 2003}
\begin{document}

\title{HST/WFPC2 Study of the Trapezium Cluster: the Influence of Circumstellar Disks on the 
Initial Mass Function}

\author{M. Robberto\altaffilmark{1}, J. Song\altaffilmark{2}, G. Mora Carrillo\altaffilmark{3}, 
S. V. W. Beckwith, R. B. Makidon, and N.~Panagia\altaffilmark{1}}
\affil{Space Telescope Science Institute, 3700 San Martin Drive, Baltimore, MD 21218}
\altaffiltext{1}{Affiliated with the Space Telescope Division of the European Space Agency, ESTEC, Noordwijk, the Netherlands} 
\altaffiltext{2}{current address: Departement of Astronomy, University of Illinois at Urbana-Champaign, USA}
\altaffiltext{3}{current address: Instituto de Astrof\'{\i}sica de Canarias, La Laguna, Tenerife, Spain}

\begin{abstract}
We have performed the first measures of mass accretion rates in
the core of the Orion Nebula Cluster. Four adjacent fields centered on the 
Trapezium stars have been imaged in the U- and B-bands using the Wide Field Planetary Camera~2
on board the {\sl Hubble Space Telescope}. We obtained photometry for 91 stars
in the U-band (F336W) and 71 stars in the B-band (F439W). The \WFPCTwo\ archive
was also searched to obtain complementary V-band (F547M) and I-band (F791W) photometry.
In this paper we focus our attention on a group of 40 stars with known spectral types
and complete UBVI \WFPCTwo\ photometry. 
We locate each star on the HR diagram considering both the standard ISM reddening law
with $R_V=3.1$ and the ``anomalous'' reddening law with $R_V=5.5$ more appropriate
for the Orion Nebula. Then we derive the stellar masses and ages by comparing with the evolutionary
tracks and isochrones calculated by D'Antona \& Mazzitelli and Palla \& Stahler.
Approximately three quarters of the sources show excess luminosity 
in the U-band, that we attribute to mass accretion.
The known correlation between the U-band excess and the total accretion luminosity, recalibrated for 
our photometric system, allows us to estimate the accretion rates, which are all found to be in the 
range $10^{-8}-10^{-12}$M$_\odot$~yr$^{-1}$.
For stars older than 1~Myr there is some evidence of a relation between mass accretion rates and stellar age.
Overall, mass accretion rates appear lower than those measured by other authors in the Orion flanking fields or 
in Taurus-Auriga. Mass accretion rates remain low even in the vicinity of the $10^{-5}$~\Msunyr\, birth~line 
of Palla \& Stahler, suggesting that in the core of the Trapezium cluster disk accretion has been recently depressed by an external mechanism. We suggest that the UV radiation generated by the Trapezium OB stars, 
responsible for the disk evaporation, may also cause the drop of the mass accretion rate. In this scenario, low-mass stars may terminate their pre-main sequence evolution with masses lower than those they would have reached if disk accretion could have proceeded undisturbed until the final disk consumption. In OB associations the low-mass end of the Initial Mass Function  may therefore be affected by the rapid evolution of the most massive cluster's stars, causing a  a surplus of ``accretion aborted" very low-mass stars and brown dwarfs, and a deficit of intermediate mass stars. This trend is in agreement with recent observations of the IMF in the Trapezium cluster.
\end{abstract}

\section{Introduction}
One of the most relevant processes in the early stellar evolution is the interaction
between young pre-main-sequence stars and their circumstellar disks. With the
exception of the very initial phases of star formation, during which protostars 
may directly accrete low angular momentum material radially infalling from their 
parental cloud, most of the stellar mass build-up occurs through disk accretion (e.g.\ Tereby, Shu, \& Cassen 1984). Disk accretion regulates the final stellar mass removing angular 
momentum via viscous dissipation and feeding conspicuous mass loss through collimated
outflows (e.g.\ Hartmann 1998).
There is evidence that the disk accretion process becomes less and less significant 
as the young star ages (Hartmann et~al.\ 1998), and, for low mass stars ($M<1M_\odot$), 
accretion must terminate within the disk lifetime (6--10~Myr, Haisch et~al.\ 2001; Skrutskie et~al.\ 1990), 
well before reaching the main sequence. However,
a quantitative understanding of the evolution of the accretion process is still missing.
In principle, mass accretion can be studied in a variety of ways. 
The infrared disk emission produced by
internal viscous dissipation is potentially an important diagnostic tool, but in most disks direct 
irradiation of the surface from the central star is a stronger heat source
(Kenyon \& Hartmann 1987). A less ambiguous indicator is the recombination radiation produced 
when the disk material falls onto the stellar surface (Bertout 1989), through a boundary layer or, more probably, through accretion columns in the stellar magnetosphere
(K\"onigl 1991; Camenzind 1990, Shu et~al.\ 1994). The hot gas continuum adds up to the stellar continuum and 
alters the depths of photospheric lines, the so-called ``veiling.'' 
From the amount by which stellar absorption features are veiled, one can 
make quantitative estimates of the relative emission of star and hot continuum, 
obtaining the integrated accretion luminosity and, therefore, the mass accretion rate
(Edwards et~al.\ 1987; Hartigan et~al.\ 1991; Bathala et~al.\ 1996; Hartigan, Edwards \& Ghandour 1995; Gullbring et~al.\ 1998). The hot continuum emission becomes easily detectable 
shortward of the Balmer discontinuity as a characteristic ultraviolet excess (Kuhi 1974). 
Gullbring et~al.\ (1998) have shown that the excess radiation in the Johnson U-band 
is related to the accretion luminosity derived from veiling measures.
Calvet \& Gullbring (1998) have reproduced the relation within the framework of the  magnetically driven accretion column model, showing that the correlation does not depend on the
main stellar parameters (mass, effective temperature) at least for spectral types
in the range from M3 to K5. 
The UV excess has been used by Gullbring et~al.\ (1998, 2000) to determine the mass
accretion of stars in the Taurus-Auriga association, by Hartmann et~al.\ (1998) in Chamaeleon-I, 
and by Rebull et~al.\ (2000) on the largely unexplored outer regions of the Orion Nebula. 

This paper presents new \HST\ photometry of 
the Trapezium cluster, the nearest region of ongoing high-mass star formation.
Located in the core of the Orion Nebula, this cluster contains at least 3000 members
between $\sim 40$~M$_\odot$ and $\sim 10$~M$_J$ (Lada \& Lada, 2003). The availability of spectral type classification for $\approx 1000$ stars (Hillenbrand 1997) provides a unique opportunity to build the
largest database of accretion luminosities and mass accretion rate.
In this work we present the results obtained for a sample of 40 isolated point sources with known spectral types 
and accurate \HST\ photometry. The observations are discussed in Section~2, whereas in Section~3 we illustrate the criteria leading
to the sample selection. Our data analysis strategy is presented in Section~4. In Section~5 we present the results, in particular the location of our sources on the HR diagram and the derived mass accretion rates. Finally, in Section~6 we discuss our findings. 
Whereas our original goal was to build the first reliable map of the mass accretion rates vs. stellar mass and age,
we find evidence of a link between the mass accretion rates and the extreme environmental conditions of the cluster core. This may have profound implications on the final stellar masses, affecting the Initial Mass Function (IMF) of the Cluster's core, especially at the low mass end.

\section{Observations and data reduction}
\subsection{HST Data sets}
Our observations were made in March 2001 using the Wide Field Planetary Camera~2\footnote{updated version of \WFPCTwo \, Instrument Handbook for Cycle 12 available at www.stsci.edu/instruments/wfpc2/Wfpc2\_hand/wfpc2\_handbook.html/} on board \HST. Four fields centered 
in the immediate surroundings of the Trapezium stars were imaged with the F336W and 
F439W filters. The F336W filter has a long cutoff wavelength at $\lambda \simeq 3600$~\AA, shortward of the Balmer discontinuity,  and is therefore similar to the Str\"omgren {\sl u} filter. The F336W filter is affected by red-leak, and in Section~4.1 we discuss how this has been taken into account.
The F439W filter closely matches the standard Johnson-B passband.
For each field, we obtained four 400 second exposures with the F336W filter and two 
180 second exposures with the F439W filter. Short exposures of 1 and 30 second were also taken to measure the bright stars.

We complement these observations by using archival \WFPCTwo~ images of the Trapezium region 
taken with F547M and F791W broad-band filters, which are close to the standard Johnson~V and Cousins~I passbands. 
The largest number of counterparts to our sources is found in the data set of GO proposal 6666 (PI Stauffer). Ten
fields were observed in November and December 1998, taking on each field two 300~second long exposures with the F791W filter, and two 500~second exposures in F547M, filter, complemented by
a 60 second exposure per filter.
Other data come from observations made in March 1995 for GO proposal 5469 
(PI Bally), where four fields were imaged in the F547M filter, with nine 40~second exposures plus five 
180~second exposure for one field, and three 30~second exposures on the remaining three. Finally, for GO proposal 6603 (PI Bally), four fields were observed in the F547M filter, with a total of three 60~second exposures and two 30~second exposures.
The first \HST\ observations of the Orion Nebula were made in 1991 
(GO proposal 2595, PI Stauffer), using the Planetary Camera channel of the Wide Field Camera (\WFC). At that time the \HST\ was still affected by spherical aberration. These data, published by Prosser et~al.\ (1994), 
have been later included in  main photometric catalog of the Orion cluster (Hillenbrand 1997), 
assembled assigning the highest priority to the observations made with the finest pixel sampling.  We shall use them for those sources lacking more recent WFPC2 photometry.

\subsection{Data reduction}
The \HST\ pipeline produces data with  bias, dark current, and flat-field 
corrections applied. Using the stellar photometry program, HSTphot\footnote{HSTphot manual available at  www.noao.edu/staff/dolphin/hstphot/}, specifically designed 
for use with \HST/\WFPCTwo~ images, we removed bad columns, cosmic-rays, and hot pixels.
To avoid objects whose spectral type classification may be affected by non-stellar emission, 
we confine our study to single, well isolated stars, and exclude sources that appear close pairs, surrounded by photoionized envelopes, and in general resolved. It must be remarked, however, that even the most prominent dark silhouette disks in Orion can only be reveled in narrow band images, since in broad-band filters they are embedded under the wings of the Point Spread Function of the central star. There is abundant evidence from IR data that disks are present around at lest $\simeq 80$\% of the source in our fields (Hillenbrand \& Carpenter 2000, Lada et~al.\ 2000), and our sample is not biased against sources with circumstellar disks.
Astrometry and photometry have been performed on individual CCD images.  
We obtained coordinates for all stellar sources by using the IRAF/STSDAS task 
METRIC, which corrects for geometric distortion of \WFPCTwo.  According to the 
\WFPCTwo~manual, the final relative positions are expected to be accurate to better than 5~mas
for targets contained within the same CCD chip, and $0\farcs1$ for targets on different chips. 
We used {\sl daophot} for multi-aperture photometry and {\sl hstphot} for psf photometry, obtaining
consistent results. The {\sl hstphot} data have been generally used for the analysis. The 
excellent quality of the residual images after PSF subtraction confirms that spurious contamination from the uniform nebular background has been virtually eliminated. The photometric zero points for each chip, filter, and gain setting were taken from Dolphin,\footnote{\WFPCTwo\ calibration data available at www.noao.edu/staff/dolphin/wfpc2\_calib/, updated on May 31, 2002} relative to
an aperture size of $0\farcs5$ (i.e. 5 pixels for the Wide Field frames and 11 pixels for the Planetary Camera frame) in radius. They are known with an accuracy of $\simeq 0.005^m$. 

\section{The sample}
On the basis of the criteria described in the previous section, we identify 91 sources in the F336W images and 71 sources in the F439W images. For 49 sources we have both magnitudes, and for 40 of them we also have a spectral classification from Hillenbrand (1997). In this paper we concentrate our attention on these 40 objects. In Table~2 we present their photometry in the four \WFPCTwo\ filters.
Since most of the stars with known spectral types are relatively bright,
they appear saturated in the archival V- and I-band \WFPCTwo\ images. For this reason, only 26 stars 
have \WFPCTwo\ photometry in all four UBVI-bands. For the remaining objects with incomplete or absent \WFPCTwo\ V- and I-band photometry we rely on the catalog of Hillenbrand (1997). 

The relevance of stellar variability can be addressed using the multiple F547M archive images taken at different epochs. The last two columns of Table~1 report the number of F547M observations and the scatter of their values. In Figure~1 we plot the individual observations, showing for each star the scatter of all available measures vs.\ their mean value. The cross symbols represent the Hillenbrand (1997) V-band magnitudes, 
that for the brightest sources are the only available data. For 36 stars we have two, or more, measures, and the standard deviations  range from a few hundredth of magnitudes to more than half a magnitude. Weighting each standard deviation by the corresponding number of measures, we obtain an average $\overline{\sigma}$(F547M)$=0.211^m$. In 2/3 of the cases (21 stars) the peak-to-peak variability is larger than $0.2^m$.
Herbst et~al.\ (2002) have recently presented the results of their study of stellar variability in the Orion Nebula cluster, done in a filter close to our F791W passband. Their study indicates that essentially all stars brighter than $I_c\simeq 16^m$, i.e.\ with photometric accuracy better than $0.01^m$, are variable. In particular, 
about half of the stars show peak-to-peak variations larger than $\sim 0.2^m$, in excellent agreement with our ratio.
Variability directly affects the derived stellar luminosity and $V-I$ color excess, since the measures in the F791W and F547M filters are in general not simultaneous. The resulting uncertainty on the extinction affects both the total and accretion luminosities. 


The V-magnitudes presented in Table~1 are the averages of all available measures, including the data of Hillenbrand (1997). The associated errors, on the other hand, are the average errors of the F547M measures made with \WFPCTwo. 
The errors are therefore indicative of our measure accuracy and not of the uncertainty on the ``true'' stellar flux. 
Figure~2 and 3 show the spectral energy distributions, together with the model results presented in the forthcoming sections.
The near-IR photometry plotted for completeness has been taken from 
Hillenbrand et al. (1998) for the J-band, and Hillenbrand \& Carpenter (2000) for the H- and K-bands.




\section{Analysis}
To obtain the mass accretion rates, we adopt the following strategy. First, for each star we compare our V- and I-band observations with the reddened spectral energy distributions of template stars, deriving the extinction towards each individual star, and the bolometric luminosity. Then, we correct the observed flux in the U-band
for the extinction, and compare it with the flux expected from a template star at the distance of the Orion Nebula. When an excess is measured, we attribute it to accretion. We then use our adaptation of the Gullbring 
et~al.\ (1998) formula to derive the total accretion luminosity. The mass accretion rate can be estimated if both the radius and stellar mass are known. The stellar mass requires the assumption of an evolutionary model. In the following subsections we detail each individual step.

\subsection{Synthetic Photometry}
To calculate the magnitudes and colors of our template stars, we adopt the  synthetic 
photometry package {\sl iraf/synphot}.\footnote{{\sl Synphot} 
manual available at www.stsci.edu/instruments/observatory/synphot.html} In addition to the four \HST\ filters, we consider Johnson~V, Cousins~I, and
Bessell JHK filters. For most of the following discussion we denote as UBVI filters those from \HST /WFPC2. We explicitely specify when referring to the Johnson-Cousins passbands.
{\sl Synphot} is also used to obtain zero magnitude fluxes from the
spectrum of Vega, which is also part of the {\sl synphot} library for  spectral calibration.
The zero-magnitude fluxes in our four \HST\ UBVI filters are, respectively: ZP(F336W)=$3.32 \times 
10^{-9}$~ergs~s$^{-1}$~cm$^{-2}$~\AA$^{-1}$; ZP(F439W)=$6.65 \times 10^{-9}$~ergs~s$^{-1}$~cm$^{-2}$~\AA$^{-1}$; ZP(F547M)=$3.55 \times 10^{-9}$~ergs~s$^{-1}$~cm$^{-2}$~\AA$^{-1}$; ZP(F791W)=$1.18 \times 10^{-9}$ ergs~s$^{-1}$~cm$^{-2}$~\AA$^{-1}$. 

The template spectra are taken from the Bruzual-Persson-Gunn-Stryker (BPGS) Spectral Atlas, which is an extension of the Gunn-Stryker (1983) optical atlas ranging from the UV to the near IR. The catalog 
contains several stars belonging to the young Praesepe and Hyades clusters that 
are representative of the stellar photosphere of our pre-main sequence objects. We select in particular the nine stars listed in Table~2. 
In order to have our templates with extinction $A_v=0$, we derive their effective temperatures from the $V-I_c$
colors using the relation provided by Hillenbrand (1997) in Appendix~C. The spectral types, alos derived using the
scale adopted by Hillenbrand (1997) in Table~1, closely match those found in the most recent literature, with the exception of BPGS59=BD+38 2457. This source has an only an historical spectral type K8, instead of the K4 value we found. In the absence of recent data, we attribute this discrepancy to an error in the original spectral classification. 
Since for M-type stars the color-temperature relations are quite uncertain, for stars later than M0 we use
the more recent scale of Reid and Hawley (2000).
For the effective temperature vs. $V-I_c$ color we derive the best fit relation
\begin{equation}
\log T_\mathrm{eff}=816.31\times(V-I_c)^3-8520.9\times(V-I_c)^2+29629\times(V-I_c)-34317 .
\end{equation} 
To locate the target stars in the HR diagram, we will apply the bolometric correction to the $V$ magnitude, $BC_V$.  In general we adopt the $BC_V$ vs.\ $\log T_\mathrm{eff}$ expression also provided in Appendix~C of Hillenbrand (1997), except for M-type stars where we use the relation
\begin{equation}
BC_V=0.27-0.604\times(V-I_c)-0.125\times(V-I_c)^2
\end{equation}
from Reid \& Hawley (2000). 

To obtain the colors and bolometric corrections of intermediate spectral types
we interpolate between the template results using the effective temperature as a free parameter.
Five target stars turn out to have effective temperatures lower than that of our coldest 
template, Gl~15B. We therefore extrapolate their colors using a second order fit to the
colors of our four coldest templates. 


\subsection{Red-leak in the F336W filter}
It is crucial to use a reliable model of the \HST\ photometric system, especially for the F336W filter which 
is affected by red-leak  (at 7500~\AA, it has 1\% of the peak transmission at 3500~\AA). 
If the red-leak is ignored, it will lead to an 
overestimate of the observed U-band excess for stars with the reddest spectral energy distributions. 
This is shown in Figure~4, where we plot for the five reddest template stars the color index F336W-F439W in function of the extinction $A_V$, both for the $R_V=3.1$ and $R_V=5.5$ reddening laws (these two values are discussed in the next Section). 
While one would expect to have a redder color when the extinction increases, i.e.\ curves rising to higher positive values, there is an anomalous ``bluing'' of the color index with the extinction (curves turn down). This behavior is well understood for \WFPCTwo: according to the instrument  handbook, the synphot predictions match the observations within 0.1\% for all template stars. 
In general, however, the corrections to the photometry for the red leak produced with these models will depend on the accuracy of the temperature and extinction estimates. These uncertainties are dominated by stellar variability, and will be discussed in Section~5.1. Figure~4 indicates that the uncertainty on the reddening law can also play some significant role. 


\subsection{Reddening}
The {\sl synphot} task is also used to calculate the colors of the template stars in the presence of reddening.
It is known that the reddening law in the direction of the Orion nebula is peculiar, with 
$R_V=A_V/E(B-V)\simeq 5$, instead of the usual $R_V=3.1$ (Baade \& Minkowski 1937, 
Johnson 1967). A higher ratio of total to selective extinction means that the dust extinction
is relatively grayer, as expected for grains larger than
those producing the standard interstellar reddening. It also means that a higher 
extinction is needed to produce the observed color excess. This effect increases 
the estimates of intrinsic stellar luminosity.

To treat the two reddening laws in a homogeneous way, we use the Fitzpatrick (1999) prescription. 
In particular, for the standard interstellar reddening law we adopt the tabulated curve for $R_V=3.1$. 
The reddening law for Orion is calculated using the IDL procedure FMRcurve.pro, also provided by Fitzpatrick (1999), with the parameters matching those in Table~5 of Fitzpartick \& Massa (1990) for $\theta^1$Ori~C (HD37022), i.e.\ $\lambda_0^{-1}=4.635$, $\gamma=0.846$, $c_1=1.251$, $c_2=0.033$, and $c_4=0.186$. The parameter $R$, which in the original Fitzpatrick's procedure is an input variable, is now calculated from the value of $c_2$, providing $R=5.504$.  
Once the reddening laws have been obtained, we attenuate the stellar templates 
by various amounts of extinction ranging between $A_V=0.1^m$ and $A_V\simeq6^m$. 
To estimate the extinction properly, one must take into account also the possible contribution of the accretion spectrum,  as discussed in the next subsection.

\subsection{Mass accretion and the Gullbring relation in the HST/WFPC2 F336W filter}
The relation of Gullbring et~al.\ (1998) (``Gullbring relation'' hereafter):
\begin{equation}
\log\left({L_\mathrm{accretion}\over L_\odot}\right)=1.09^{+0.04}_{-0.018} \cdot\log\left({L_{U_J, \mathrm{excess}}\over L_\odot}\right)+0.98^{+0.02}_{-0.07},
\end{equation}
linking the Johnson U-band excess, $L_{U_J,\mathrm{excess}}$ to the total accretion luminosity, $L_\mathrm{accretion}$, 
is based on observational data and is valid for the Johnson-U filter only. 
Our F336W filter is better suited to this type of study than the Johnson-U 
filter, as the F336W bandpass is entirely on the blue side of the Balmer jump, 
whereas the Johnson-U filter runs across it. This is an advantage, since, in 
general, the photospheric emission of young stars drops or remains constant 
shortward of the Balmer discontinuity, while the ionized emission from the accretion material increases. 
In particular, whereas in the Johnson U-band the UV excess is higher for late 
spectral type stars than for early-type stars, due to partial cancelation of the emission and
absorption from the ionized gas and hot photospheres, in our F336W filter the excess 
is expected to remain consistently high and increasing with the stellar temperature. 
The Gullbring relation was reproduced by Calvet \& Gullbring (1998) on the basis of magnetospheric infall models (Uchida \& Shibata 1984, Camenzind 1990, K\"onigl 1991). They assumed that the material, migrating from the disk to the stellar surface through accretion columns, produces shock emission that can be calculated with a 1-d model. In the wavelength range shortward of 
the Balmer discontinuity the spectrum is dominated by optically thin emission from the pre-shock gas and 
from the attenuated post-shock regions, whereas the Paschen and Bracket continua are mostly produced by the optically thick emission from the heated photosphere below the shock. The  Gullbring relation predicted by this model has a slightly shallower slope than the observed one, with no evidence of dependence on the stellar spectral types of the underlying photosphere, at least in the range from K5 to M3. 

Since our main goal is the calibration of the Gullbring relation to the F336W filter, we are mostly interested in the local behavior of the spectrum in the vicinity of the Balmer jump. Thus, we assume for simplicity that the excess is produced by a uniform and isothermal slab of ionized gas. Although a single slab cannot reproduce the entire accretion spectrum (Calvet \& Gullbring 1998), it allows us to calculate a grid of models easily and to select among them those matching the original Gullbring relation within the errors.  Our assumption is that the sub-set of models that reproduce the observed relation between $L_{U_J,\mathrm{excess}}$ and 
$L_\mathrm{accretion}$ will also provide a reliable relation between $L_{F336W,\mathrm{excess}}$ and $L_\mathrm{accretion}$. In other words, we  cross calibrate our slab models with the observational data in the region of the Balmer discontinuity.

To generate the slab models, we use the photoionization code CLOUDY (Ferland 1996\footnote{Updated on 2002 program and documentation for Beta 4 of Cloudy 96 available at http://nimbus.pa.uky.edu/cloudy/cloudy\_96b4.html}) with a range of densities and temperatures matching those considered in previous studies (Batalha, Lopes, \& Batalha 2001; Calvet \& Gullbring 1998). In particular, the densities are in the range $\log n=14.6-15.2$ and the temperatures in the range $\log T=3.88-3.92$. We consider the typical composition of HII regions as a baseline, but also explore values of 10 times higher or lower metallicity. 
The slab has a thickness of $10^{7.3}$\,cm, and an excitation source of 10~L$_\odot$ with a blackbody spectrum of $T_\mathrm{eff}=10,000$\,K, at a distance $d=10^{16}$~cm from the slab. 

Each synthetic spectrum is convolved with both the Johnson-U filter and the F336W filter. The spectra matching the Gullbring relation within the errors are selected to generate the new relation between the accretion luminosity in the F336W filter and the total accretion luminosity. The revised Gullbring relation for the F336W filter is
\begin{equation}
\log\left({L_\mathrm{accretion}\over L_{\odot}}\right)=(1.16 \pm 0.04) \cdot \log\left({L_{F336W,\mathrm{excess}} \over L_{\odot}}\right)+(1.24 \pm 0.10).
\end{equation}
The higher intercept of this relation (1.24 vs.\ 0.98) indicates that our F336W bandpass, narrower than the Johnson~U bandpass, includes a smaller fraction of the accretion luminosity. On the other hand, the higher slope (1.16 vs.\ 1.09) indicates that the F336W bandpass is a more sensitive indicator of the accretion luminosity. 

The ionized gas responsible for the UV excess shortward of the Balmer discontinuity also affects the entire spectrum. Hartigan et~al.\ (1991) have shown that the accretion continuum may contribute significantly to the flux in the V and R~passbands, affecting the extinction estimates by $\simeq 50$\%. The emission in the Paschen continuum, in particular, may increase the measured flux in the I-band appreciably. 
To account for this contribution, we consider a representative slab model at $T=7600$~K and repeat the procedure followed for the stellar templates, using {\sl synphot} to predict the flux in all broad-band filters with differing amounts of reddening. Moving away from the Balmer discontinuity, our single slab models become less reliable, but still adequate to our purpose.

\subsection{Model fit and accretion luminosity}

To calculate the extinction, we use an iterative procedure. First, we determine $A_V$ by comparing
the observed vs. template colors. Then we add our representative accretion spectrum, with the same amount of
extinction just obtained for the pure photosphere, 
and match the observed U-band magnitude by adjusting a multiplicative constant that can be interpreted as a filling factor. A new extinction estimate is then obtained by comparing the observed 
data with the star+accretion spectrum, and the process repeated. With respect to the study of Hillenbrand (1997),
which was based on the V-I color and assumed the standard $R_V=3.1$ reddening law, we have the advantage of the extra B-band.  A best fit to the BVI photometry provides a more reliable estimate of the extinction 
than the V-I color only, and possibly a hint on the best reddening law. On the other hand,
the B-band data play in our case a special role. Since the B-band and the U-band images have been taken almost simultaneously, we expect the U-B color to be quite independent on source variability, being possibly affected only by flaring activity in the accretion process, but not by variability modulated by stellar rotation. In other words, to estimate the UV excess it is convenient
to constrain the model to match exactly the observed B-band photometry. Our strategy is therefore 
to obtain the extinction from the best-fit to the observed BVI colors and the average stellar luminosity, 
and then normalize it to pass through the B-band point, providing in this way a most reliable estimate of the UV excess at the time of the observations. Technically, we use a least
square estimator. To calculate the  $\chi^2$, we weight each deviation squared with the variances obtained by adding quadratically three contributions:
the typical errors of the original measures in each band (reported in Table~1), the zero point 
errors quoted by Dolphin, and the standard deviations $\sigma_V(F547M)$ listed in the last column of Table~1, 
assumed to be representative of the stellar variability. 
In the four cases with only one available observation in the F547M filter, 
we assume that variability is present at the level of the mean standard deviation of the sample, $\overline{\sigma}=0.211^m$. This procedure is carried out for both reddening laws. 

Once the extinction is known, the total stellar luminosity is obtained through the formula
\begin{eqnarray}
\log\left({L_*\over L_{\odot}}\right) = 0.4(M_{\odot,\mathrm{bol}} - M_\mathrm{tot,bol}) = 0.4(4.75 - V - BC_V + A_V + DM),
\end{eqnarray}
where we assume $M_{\odot,\mathrm{bol}}=4.75^m$ for the solar absolute magnitude and a distance module to the Orion cluster, $DM=8.36^m$, corresponding to a distance of 470~pc. 

The accretion luminosity in the U-band,
$$
L_{U,\mathrm{excess}}\equiv L_{U,\mathrm{observation}} - L_{U,\mathrm{expected}},
$$
refers to the luminosities corresponding to the dereddened magnitudes.
The Gullbring relation for the \HST\ U-band derived in the previous section (\S4.4)
provides the total accretion luminosity, $L_\mathrm{accretion}$. 
We estimate the mass accretion rates by using the standard relation between $\dot{M}$ and $L_\mathrm{accretion}$, 
\begin{equation}
\dot{M} = {{L_\mathrm{accretion} R_\star}\over {0.8 GM_\star}},
\end{equation}
where the stellar radius, $R_\star$, is obtained from the standard $L_\star=4\pi R_\star\sigma T_\star^4$ relation with $\sigma=5.67\times10^{-5}$~erg\,cm$^{-2}$\,s$^{-1}$\,K$^{-4}$, and the stellar mass, $M_\star$ is obtained from theoretical models of the pre-main-sequence evolution (\S5.2). The factor of 0.8 accounts for the fact that the material is assumed to be
freely falling along the magnetosphere from a corotation radius $R_{mag}\sim 5R_\star$ (Gullbring et~al.\ 1998). 

\section{Results}
\subsection{Spectral Energy Distributions}
Figures~2 and 3 show the model fits to the observed magnitudes 
for the $R_V=3.1$ and  $R_V=5.5$ reddening laws, respectively.
For each star, we plot the measured magnitudes, the model  spectrum (star+accretion)
matching the BVI photometry (dotted line), the same model normalized to pass through the B-band photometry for the estimate of the U-band excess (solid line), and, when relevant, the accretion spectrum (dashed line). 
In Table~3 and Table~4 we list the corresponding results for the stellar luminosity (columns~3), the extinction (column~4), the stellar radius (column~5), and the accretion luminosity (column~6), again for both reddening laws. The effective temperatures (column~2) have been taken from the Hillenbrand 1997 scale.

On the highly compressed scale of Figures~2 and 3, the dotted lines nicely match the V- and I-band
data. The displacement with respect to the solid lines, i.e.\ the deviation of the B-band with respect to the best fit, visualizes the range of variability between the average magnitude and the magnitude at epoch of our UV observations. In almost all cases the U-band flux appears strongly in excess with respect to the ideal
reddened black-body curve, but this is largely due to the red-leak. The real accretion luminosity is represented by the distance between the U-band point and the solid line. When this is large,  the standard accretion spectrum, normalized to account for the U-band flux, becomes evident at the bottom of the figure. 
The difference between the values of $A_V$ derived from the two reddening laws is of the order of 15\%.
The stellar luminosities are also affected in a similar way, due to Equation~5. 

To assess the reliability of our model fits we can refer to the $\chi^2$ parameter that is used to minimize
the discrepancy between the model fit and the observed B, V, and I magnitudes, as described earlier.
If one assumes that the deviations are approximately Gaussian, i.e.\ with zero mean and
independent, then the distribution of the $\chi^2$ values should follow the theoretical reduced $\chi_\nu^2$ distribution, where $\nu$ is the number of degrees of freedom. In our case, it is $\nu=3$ since for each star we typically have five bands (the \WFPCTwo~ BVI and the VI data from Hillenbrand (1997) minus two model parameters (reddening and normalization). Poor fits provide highly improbable $\chi^2$ values, above some defined critical threshold. With 40 sources, we can set the critical threshold at the 99\% probability, $\chi_3^2(0.99)=11.3$. Above this value we should find 1\% (i.e.\ none) of our stars, whereas we have approximately 
50\% of our sources, no matter which reddening law we adopt. According to the standard prescriptions, for these objects we should reject the hypothesis at the basis of the model fit, e.g.\ the quoted spectral types, or the adopted reddening law, or both. On the other hand, if systematic contributions to the noise are underestimated, a $\chi^2$ behind the critical threshold may force wrong decisions. In our dataset, stellar variability is a prime suspect for loose fits, since our sparse sampling in time certainly underestimates the variability range.

When the measured $\chi^2>11.3$ and $\sigma(F547M)<\overline{\sigma}=0.211^m$, we make the assumption that variability has been underestimated. If we recalculate the $\chi^2$ forcing in these cases $\sigma(F547M)=\overline{\sigma}$, nearly half of the sources above the critical threshold can be recovered. The resulting $\chi^2$ distributions, shown in Figure~5 together with the theorical $\chi_3^2$ distribution, show that the number of outliers above $\chi^2\simeq6$ (where the probability falls to approximately 10\%) is still too large. Figure~5 clearly shows the limitations intrinsic to the dataset at our disposal. In particular, all sources with an anomalously high value of $\chi^2$ must be considered with great caution. For this reason, in the following we shall 
distinguish between a main sample, composed by sources with $\chi^2<11.3$ that are 
expected to have a good fit, and a secondary sample with $\chi^2>11.3$, having a less reliable fit. 
A Kolmogorov-Smirnov test on the two distributions of the observed $\chi^2$ presented in Figure~5 indicates that the hypothesis that both are drawn from the same population must be accepted with a very high degree of confidence. In other words, the available data do not allow to select statistically one reddening law or the other. Since the reddening law affects the estimates of the mass accretion rate, we will continue to carry out our analysis for both laws.

A last remark concerns the excess UV luminosity. Only in 7 cases ($R_V=3.1$) or 9 cases ($R_V=5.5$)
there is no evidence of a UV  excess. The lack of excess may be due to short term variability, related e.g.\ to flaring, which is not properly sampled in time-scale of our UB observations.  In Figure~6 we plot the histogram of the U-band excess luminosity for  the two different reddening laws. The strong asymmetry to the positive side, even at the lowest signal levels, indicates that the F336W excess is real. Figure~2 indicates that only a systematic underestimate of the extinction could artificially produce an excess emission in the UV, especially for the most reddened stars. Given that the main uncertainty in our best fit is caused by random stellar variability, 
we tend to rule out major systematic errors. Since the ratio between the accretion luminosity and the total stellar luminosity, corresponding to the veiling factor of the stellar absorption lines, is generally small, the positions of the stars in the HR diagram remain almost unaffected by the removal of the accretion contribution.



\newpage
\subsection{The HR diagram}
The location of our 40 stars in the HR diagram is shown in Figure~7. The four plots have been obtained assuming the $R_V=3.1$ (left column) and $R_V=5.5$ (right column) reddening laws, and two different sets of theoretical models. On the top row we use the latest version of the isochrones and evolutionary tracks of D'Antona \& Mazzitelli.\footnote{available at http://www.mporzio.astro.it/~dantona/} (1998, hereafter DM98), whereas on the bottom row we use the results of Palla \& Stahler (1999, hereafter PSF99) The different symbols
identify the sources according to their reliability class: filled dots are used for the more reliable sources of the main sample, small dots with circles are used for the remaining more uncertain sources. On each figure,
the two diagrams located at the bottom at $\log(T_{\rm eff})=3.694$ (K2 spectral type) and at the top at $\log(T_{\rm eff})=3.502$ (M3.5 spectral type) represent how different types of errors change the position of a stars in the diagram. Namely, 1)~the vertical bar represents the $\pm 1\sigma$ error associated to the average 
photometric uncertainty $\Delta V=0.211^m$, that we attributed to stellar variability;
2)~the horizontal bar represents the error on the effective temperature, reported by Hillenbrand (1997) to be equal to $\pm~1$ subclass (corresponding to $\Delta\log T_{\rm eff}\approx \pm0.011$) for stars K7 (i.e.\  $\log T_{\rm eff}=3.602$) and later, and $\pm1/2$~class ($\Delta\log T_{\rm eff}\approx \pm0.05$) for earlier stars; 3)~the diagonal bar represents the error in the bolometric correction that is introduced
when a inaccurate spectral type is assumed. Since for our late spectral types the bolometric correction increases (becomes more negative) with the spectral type, a star of a given magnitude will be more luminous if a lower temperature is attributed to it. A lower temperature will also make the spectral type redder,  thus reducing the estimated extinction; this compensates for the error on the bolometric correction by an amount that depends on the extinction. In the plot we refer to the  case without extinction. 


Using theoretical evolutionary models, one can interpolate between evolutionary tracks and isochrones to derive the mass and the age of each individual star. 
In our case, we are mostly interested in the mass value as it enters in Equation~6 for the estimate of the mass accretion rate.\footnote{In Section~6 we will compare the mass accretion rates in Orion with those in Taurus, estimated by Hartmann et~al.\ (1998) using the DM98 family of models.}
The results, also listed in Table~3 and 4, 
are strongly model dependent for a variety of reasons as discussed, e.g., by
Baraffe et~al.\ (2002). The two particular models we have considered are characterized by different choices in the initial conditions (stellar radius), but in addition other physical processes, like chemical composition, convection efficiency, surface gravity, and even the accretion rates that we are trying to measure, contribute to provide discrepant results. According to Baraffe et~al.\ (2002), at present there is no reliable theoretical model for ages $<10^6$~yr. Even if absolute ages and masses cannot be unambiguously determined, the comparison of the results obtained using different models may provide useful insights. 

Using DM98 models, masses range from  to 4.2~M$_\odot$
to less than 0.08~M$_\odot$, the H-burning limit. The values, however, 
may vary by a factor of $\simeq 2.0$ if one uses different PMS models, especially in the range
$\sim0.6 - 1~M_\odot$.
For example, source 9209 has $\log T=3.58$ and $\log(L/L_\odot)=3.078$ in the reference b) case. The 
corresponding mass is $M=0.26$~M$_\odot$ using D'Antona and Mazzitelli tracks, or $M=0.57$~M$_\odot$ 
using PS99 tracks. The stellar masses of the sources above the 0.1~Myr isochrone (for DM98) or above the
birthline (for PS99) have been estimated using the mass of a source lying on the 0.1~Myr isochrone or birthline at the same effective temperature. On the other hand, there are five sources that lie below the $0.1~M_\odot$ evolutionary track of PS99. These most interesting brown dwarf candidates fall, in the PS99 case, in our secondary sample. Given the uncertainties, 
we did not attempt to extrapolate our fitting process behind the $0.1~M_\odot$ track.
With these provisions, the average mass of our sample turns out to be
\begin{enumerate}
\item D'Antona \& Mazzitelli: $<\log M(M_\odot)>=-0.52\pm0.35$
\item Palla \& Stahler: $<\log M(M_\odot)>=-0.371\pm0.37$
\end{enumerate}
essentially regardless of reddening law. 

Concerning ages, the evolutionary tracks and isochrones of D'Antona \& Mazzitelli have been used by Hillenbrand (1997) to derive a cluster age less than $1\times10^6$~yr
for masses between 0.01 and 2.5~M$_\odot$. Using near-IR data and updated D'Antona \& Mazzitelli (1998) models, Hillenbrand \& Carpenter (2000) found that a star formation rate constant between $3\times10^4$ and $3\times10^6$ years in logarithm of the age bins closely matches the distribution of stellar ages. Luhman et~al.\  (2000) compared their HST/NICMOS near infrared observations and ground-based K-band spectra with D'Antona \& Mazzitelli (1998) models and found a median age of $4\times10^5$ years. 
PS99, on the other hand, used their models on the same dataset of Hillenbrand (1997), deriving 
a best fitting single age of $2\times 10^6$~yr, with star formation starting at low level some 
$10^7$~years ago and accelerating to the present epoch (but see Hartmann 2001 for a discussion of the systematic errors). PS99 models use deuterium burning to control the protostellar radius during the 
initial accretion phase. This sets a well defined locus (birthline) in the HR diagram. The birthline shown in 
Figure~7 has been calculated for an assumed mass accretion rate of $\dot{M}=1\times10^{-5}$~M$_\odot$yr$^{-1}$. 

Considering the logarithm of the ages, as recommended by Hartmann (2001), we obtain ages compatible with previous estimates 
based on these families of models:
\begin{enumerate}
\item DM98+$R_V$=3.1: $\log t($yr$)=5.76\pm0.90$ (average); $\log t($yr$)=6.02$ (median)

\item DM98+$R_V$=5.5: $\log t($yr$)=5.51\pm1.07$ (average); $\log t($yr$)=5.79$ (median)

\item PS98+$R_V$=3.1: $\log t($yr$)=5.99\pm0.55$ (average); $\log t($yr$)=6.07$ (median)

\item PS98+$R_V$=5.5: $\log t($yr$)=5.80\pm0.58$ (average); $\log t($yr$)=5.94$ (median)
\end{enumerate}
If the best assumptions are those that minimize
the scatter with respect to an average cluster's isochrone (Hartmann 2001), then
PS99 models seem preferable. PS99 models also predict isochrones that apparently show
a better match to the overall distribution of stars in the HR diagram. In particular, in the $R_V=5.5$ case the maximum age is consistently close to the 3~Myr isochrone, whereas the D'Antona \& Mazzitelli isochrones show a correlation between ages and mass, with more massive stars appearing younger. This trend was already noticed by Luhman et~al.\ (2000), Hillenbrand \& Carpenter (2000) and Muench et~al.\ (2002).
The age distribution in this case (Figure~8) shows some evidence of the elusive  ``pileup'' of stars at the maximum age expected when star formation is concentrated in a relatively short episode (Hartmann 2001).
The increase of the number density of stars is roughly linear with the logarithm of the age, and can be approximated by the relation $\log dN=3(\log t -4)$. 
The observed logarithmic decline actually represents an increase of the star formation rate with time, 
similar to the ``acceleration'' suggested by PS99. In this scenario, the Orion cluster can be considered as an
{\sl active} site of star formation.


\subsection{Mass Accretion Rates}

Following the method described in Section 4.5, the mass accretion rate, $\dot{M}$, can be estimated for approximately 3/4 of the stars. The values, listed in Table~3 and 4 for both evolutionary models and reddening laws, range between $\simeq 2\times10^{-8}-10^{-12}$\Msunyr. For the remaining 1/4  
of stars, $\dot{M}$ cannot be obtained either because there is no mass estimate (the star lies in the HR diagram
out of the model range) or because there is a deficit of UV flux with respect to the stellar photosphere.  
Stars of the main sample have typically mass accretion rates lower than those in the secondary sample. Moreover,  stars with UV deficit are typically part of the main sample. In conclusion,
when a star is well measured, the mass accretion rate is always low, or absent.
The average and median values for the main sample are:
\begin{enumerate}
\item
DM98 + $R_V=0.3$: average=$3.59\times10^{-9}$~\Msunyr; median=$7.4\times10^{-10}$~\Msunyr
\item
DM98 + $R_V=5.5$: average=$3.12\times10^{-9}$~\Msunyr; median=$1.0\times10^{-9}$~\Msunyr
\item
PS99 + $R_V=3.1$: average=$3.90\times10^{-9}$~\Msunyr; median=$6.0\times10^{-10}$~\Msunyr
\item
PS99 + $R_V=5.5$: average=$2.39\times10^{-9}$~\Msunyr; median=$8.6\times10^{-10}$~\Msunyr
\end{enumerate}
In Figure~9 we show the same four HR diagrams of Figure~7, using for each star a circle with diameter
proportional to the logarithm of $\dot{M}$, if measured. Thick and thin circles represent main and secondary sources, respectively. Figure~9 shows no obvious correlation between mass accretion rates and the position of the star in the HR diagram. A direct comparison of the four plots reveals how different reddening laws and PMS evolutionary models affect the estimates of $\dot{M}$. The uncertainty on the reddening law affects the measure of $\dot{M}$ through the U-band extinction, the red-leak correction and the stellar radius, derived from the stellar luminosity. Evolutionary models provide different values of the stellar mass, that directly enters in the estimate of $\dot{M}$ through Equation~6. Table~3 and 4 show that the average scatter between different estimates of accretion rates is of the order of 30\%.


Besides the reddening laws and theoretical models, there are other sources of uncertainty.
Since the Orion cluster is relatively far and compact, the relative distance of cluster members has a negligible effect on the scatter of the luminosities and mass accretion rates, assuming the cluster depth of the order of the projected cluster size, $\simeq 1$~pc. A major systematic
uncertainty is the average distance of the Cluster, which is known to within a $\sim 20\%$ error. It ranges from $d\simeq400-480$~(Warren \& Hesser 1977) to $470\pm 80$ (Genzel et~al.\ 1981).
Changing the distance from 470~pc to 430~pc decreases the stellar luminosities by a factor $\approx 20$\% 
and the mass accretion rates by a factor $\approx 28$\%, on average. Another systematic error is introduced by the assumed value for the radius at which magnetospheric infall begins, $R_\mathrm{mag}$. The value we assumed, $R_\mathrm{mag}=5$, is equal to that assumed by Hartmann et~al.\ (1998) for their study of the Taurus and Ophiucus region, but other choices are possible. Muzerolle et~al.\ (2003)
have recently assumed for instance $R_\mathrm{mag}=3$, on the basis of their H$\alpha$ emission line models. This choice would lead to mass accretion rates lower by $\simeq 20$\%. 
Concerning photometric errors, they also contribute to the uncertainties on the accretion rates. 
An errors of $\sim 0.05^m$ in the U band magnitude adds $\simeq 5$\% of uncertainly. In general, measurement errors provide a negligible uncertainty with respect to systematic effects we have discussed. On average, our results should be correct to within a factor of 2. 

\section{Discussion}
\subsection{UV excess as a tracer of mass accretion}
We have selected a sample of young ($t\lesssim3$~Myr), low mass 
($M\simeq0.1-1$~M$_\odot$) stars with evidence of mass accretion ($\dot{M}\simeq10^{-8}-10^{-12}$~M$_\odot$yr$^{-1}$) in $\simeq 75\%$ of the cases. 
The interpretation of UV excess as entirely due to mass accretion rates can be questioned. Disk photoevaporation
creates envelopes surrounded by a photoionization front facing the ionizing stars (see later). The recombination flux emitted by these low density ionization fronts should be quite different from that produced in the
shock excited high-density accretion column, but our broad-band photometry does not allow to distinguish between these two processes. To mitigate possible contamination, we have rejected from our sample all sources appearing clearly extended on a visual inspection. A second check is offered by Figure~10, where the location of our sources in the Trapezium core is shown. Each star is represented by a circle with diameter proportional to  the excess UV luminosity. If radiation by Trapezium stars was to be the main source of stellar envelope ionization,  one would expect to find a strong gradient of the UV flux with the distance from the core
of the Nebula, which is not observed. Within the area covered by our study there is no clear trend between the excess UV luminosity and the projected distance from the center of the nebula.  In any case, any contamination from the envelope would increase the UV excess, forcing us to consider the values we have found as upper limits. Since the accretion rates appear already anomalously low, this effect would just strengthen our conclusions.


The interpretation of the UV excess as actually due to the accretion process is strengthened by the strong correlation with the near IR excess, represented in Figure~11 through the K-band excess. This relation has been quite elusive in the past, but in our case
turns out to be evident. This because we can rely on accurate photospheric subtraction, rather than on color indexes. There scatter is still large, and category~1 stars with reliable spectral energy distributions show an excess stronger in the near-IR than in the UV.  One must take into account that the UV and near IR emission come from different regions, and at small distances from the star the disk orientation plays a dominant role shadowing the inner disk region. A more detailed study of the IR excess emission may lead to constrain the inner holes and reduce the uncertainties on the assumed $R_\mathrm{mag}$. In any case, the presence of substantial IR excess indicates that possible inner holes have small size, coherently with a scenario where the accretion process proceeds  across the disk down to the magnetospheric radius. It is interesting to notice that if the growth of protoplanets is traced by the development of large disk gaps, our disks appear to be in an earlier evolutionary phase.


\subsection{Low mass accretion rates}
In Figure~12 we plot the mass accretion rates derived assuming the DM98 and 
PS99 models and $R_V=3.1$ and $R_V=5.5$ reddening laws, versus the stellar ages. Black filled and open circles are used for main and secondary sources, respectively, and their size is proportional to the stellar mass. To reduce the uncertainties related to the  definition of the zero age, we have considered only stars with ages older than 0.1~Myr.


The red circles refer to the Taurus sources studied by Hartmann et~al.\ (1998) assuming $R_V=3.1$ and the DM98 tracks, and using, for most sources, the same UV excess technique adopted in our study. For their sample Hartmann et~al.\ (1998) found a decrease of the mass accretion rate with stellar age. In our case, the possibility of a correlation depends on the assumptions. A Spearman rank-order test, which being non parametric is independent on the actual values of age and accretion rate, shows that chance of a random correlation varies in the four cases according to:
\begin{enumerate}
\item
DM98 + $R_V=0.3$: P=3.7\% (main sample), P=3.6\% (all stars)
\item
DM98 + $R_V=5.5$: P=0.9\% (main sample), P=60\% (all stars)
\item
PS99 + $R_V=3.1$: P=0.02\% (main sample), P=0.09\% (all stars)
\item
PS99 + $R_V=5.5$: P=13\% (main sample), P=18\% (all stars)
\end{enumerate}
If we restrict to the more reliable sources of the main sample, evidence of correlation seems especially strong in the case of PS99 + $R_V=3.1$. In this case, a power law fit to the entire sample shows that $\dot{M}$ decreases as $\dot{M}(t)=Kt^{-\eta}$, with $\eta=1$. The scatter, however, is large, 
with mass accretion rates that differ by about two order of magnitudes at the same stellar age. For comparison,
Hartmann et~al.\ (1998) find $1.5\lesssim\eta\lesssim2.8$ for Taurus and Cha~I associations, and comparably large scatter. A more quantitative analysis of the correlation is complicated by the fact that the age and accretion rate are intrinsically correlated,  being both derived from the stellar luminosity. 

The steeper decline with time (higher $\eta$) for the Taurus sample is not the only difference with our Orion results.  Figure~12 clearly shows that mass accretion rates in Orion are systematically lower: in the previous section we have quoted average values $\simeq10^{-9}$~\Msunyr, whereas Hartmann et~al.\ (1998) find for Taurus an average of $10^{-8}$~\Msunyr. Since Hartmann et~al.\ (1998) sample contains a similar number of stars with ages comparable to our sources, we tend to reject the hypothesis that all of our stars have been observed by chance in a quiescent phase. The same mean value of Hartmann et~al.\ (1998) was also found by Rebull et~al.\ (2000) in their study of the mass accretion rates in the Orion Nebula flanking fields, four $45\arcmin\times45\arcmin$ fields centered $\sim0\fdg5$ east, west, north, and south of the core area covered by our study. Even if we consider the stars younger than 0.1~Myr, not plotted in Figure~12, the mass accretion rates remain of the order of $10^{-9}$~\Msunyr. The flatter decay curve we find in Orion therefore appears related to a recent systematic leveling off of the mass accretion rates, while one would have expected finding some significant increase at the young ages probed by our study.  PS99 have shown that the locus of the Trapezium stars in the HR diagram is nicely enveloped, at the high luminosity side, by the birthline calculated for a protostar with accretion rate of $10^{-5}$\Msunyr\ and deuterium-to-hydrogen ratio D/H=$2.5\times10^{-5}$. The locus of the birthline in the HR diagram does not depend upon the initial mass and radius of the collapsing protostellar core, but depends on the mass accretion rate: given two cores with the same mass, the one having lower mass accretion rate will be more compact. Due to the higher  density, it will ignite and deplete its deuterium content earlier and be less luminous (Stahler 1988). The $\dot{M}\simeq 10^{-5}$\Msunyr\, value of PS99 is compatible with stellar collapse studies (Masunaga, Miyama,  \& Inutsuka 1998) and provides a good agreement with the observations, not only in what concerns the birthline, but also the zero-age main-sequence locus and the evolutionary tracks (Palla \& Stahler 2001). 

Stars on the $10^{-5}$~\Msunyr\ birthline may show lower mass accretion rates.  Hartmann, Cassen, \& Kenyon (1997) have found that if the accretion occurs mainly from an equatorial disk (rather than from a spherical 
envelope as in PS99), a sudden drop in the mass accretion rate has no immediate influence on the
photospheric effective temperature and luminosity. The residual time spent on the birthline, or in its vicinity, 
can be easily estimated, since the energy available from Deuterium fusion in a core of mass, $M_c$,
is $Lt=1~L_\odot(1.5\times10^6$yr$)(M_c/M_\odot)$. A completely convective, non accreting, $M_c=0.2M_\odot$ PMS star on the $10^{-5}$~\Msunyr\ birthline will need $\approx10^5$~yr to resume contraction. This time drops by about an order of magnitude at $M_c\simeq0.5~$M$_\odot$, becoming negligible at higher masses. Therefore, in order to maintain in the vicinity of the birthline a group of stars with mass spread larger than an order of magnitude and very small mass accretion rates, one must assume that the main accretion phase has been recently terminated for all sources, independently on their ages. In principle, this can be simply due to the spontaneous exhaustion of disk material available for accretion. However, it seems rather artificial to have all disks depleted almost at the same time across the cluster. A more attractive possibility seems to attribute the drop of mass accretion to a common trigger event caused by an external agent. 

It is known that disks in Orion are photo-ablated by the UV radiation of OB stars, the ``proplyd'' phenomenon 
first found by Churchwell et~al.\ (1987) and later confirmed by spectacular \HST\ images (O'Dell, Wen, \& Hu 1993, O'Dell \& Wen 1994).  Churchwell et~al.\ (1987) estimated mass-loss rates of about $10^{-7}$~\Msunyr, whereas Henney \& O'Dell (1999) more recently have measured mass-loss rates as high as $10^{-6}$\Msunyr. Thus, when a protostar is exposed to the ionizing flux of a new-born OB star, the disk mass decreases rapidly with time. The similarity solution of Hartmann et~al.\ (1998) provides a theoretical link between the disk mass and the mass accretion rate.  In Figure~12 we show the $\dot{M}$-age relations expected  for viscous disk accretion. The solid line corresponds to the fiducial model with $\eta=1.5$, stellar mass $M_\star=0.5$~M$_\odot$, radial disk scale factor $R_1=10$~AU, outer disk temperature $T_{d2}=10$~K, viscosity parameter $\alpha=0.01$, and initial disk mass $M_d(0)=0.1$~\Msun. The dashed line refers has the same parameters except for $M_\star=0.1$~\Msun, whereas the dotted line has $M_\star=0.1$~\Msun\, and 
$M_d(0)=0.01$~\Msun. In general, the mass accretion rate is proportional to the disk mass and decreases with time.  It is important to note that this model refers to the {\sl disk} mass accretion, $\dot{M}_d$, whereas
our observations refer to the {\sl stellar} mass accretion, $\dot{M}_\star$. Mass conservation implies that part of the accreted disk mass is lost through the outflow activity at a rate $\dot{M_w}$. The standard model for magnetocentrifugally driven winds  predicts $\dot{M}_d/\dot{M}_w\approx3.5$ (Shu et~al.\ 1994), thus 
$\dot{M}_d=1.4\dot{M}_\star$, i.e.\ disk and stellar mass accretion rates are of the same order. 
In Orion, mass loss is typically traced by ``microjets'' with modest mass-loss rates, around $10^{-9}$\Msunyr\, (Bally, O'Dell, \& McCaughrean 2000). 

Since the mass loss from the disk evaporation dominates all other mass transfer mechanisms, it is tempting to speculate that the reduction of disk mass resulting from the exposure of a circumstellar disk to the UV radiation of newly born OB stars regulates the mass accretion rate through the disk, and therefore to the star. 
The similarity solution, based on assumptions like thin disk geometry and negligible external heating, suggests
a possible link: the mass accretion rate is proportional to the disk mass. The similarity solution, however, 
can hardly be directly applicable to real photoevaporated disks. In a previous paper we have shown that when the circumstellar disk becomes exposed to the UV radiation, the disk flaring angle (i.e., thickness) and temperature show a dramatic increase, depending mostly on the distance and orientation of the disk with respect to the main heating source (Robberto, Beckwith \& Panagia 2003). An increase of the outer disk temperature raises the efficiency of internal energy dissipation, supporting higher mass accretion rates. Further theoretical investigations are required to understand the interplay between disk photoevaporation and mass accretion.

\subsection{Implications for the star formation history and IMF}
If the mass accretion rates are affected by the disk photoevaporation phenomenon, one may expect to find systematic differences between the stellar population in the core of the Orion Nebula and other regions, like Taurus or the outskirts of the Orion Nebula itself, where the ionizing flux is negligible. For PMS sources still in the disk accretion phase, the sudden decrease of the mass accretion rate causes an premature end of the stellar
mass build-up. Whereas the final stellar mass is fixed in the early stages of protostar formation, through competitive collapse/fragmentation phenomena regulated by magnetic fields or supersonic turbulence, most of the stellar material is accreted from the circumstellar disks at a rate which decreases with time. There is growing evidence that low mass stars and sub-stellar mass objects share this type of evolution (Muzerolle et~al.\ 2003), possibly on longer time scales. These may especially be affected by the sudden disk evaporation, growing to a final mass lower than that they would have attained if the star formation process would have proceeded in a more quiet environment. Low mass objects, therefore, may remain "dwarfed" by the sudden disk dissipation,
resulting in a relative overabundance of low mass stars and brown dwarfs. Various authors, and in particular Luhmann et al. (2000), have recently reported an overabundance by a factor of 2 in brown dwarfs in Orion relative to Taurus. At the same time, the overabundance of low mass stars and brown dwarfs should be compensated by a depletion of intermediate mass stars. In this mass range, comparison are affected by small number statistics. Still, it is intriguing to notice that Hillenbrand (1997) found a flattening of the IMF in the core of the Trapezium cluster at masses lower the 0.6~$M_\odot$, whereas the overall Orion stellar population follows the Salpeter law down to the completeness limit of her survey, $0.1~M_\odot$.  This flattening could be the final outcome of the disk depletion, causing all stars cascading in lower mass bins. The ultimate consequence of this scenario is that the low-mass end of the initial mass function could be  modulated by the recent star formation history, namely by the formation of the OB stars in the cluster.

\section{Conclusion}
We have performed the first quantitative analysis of the mass accretion rates in the core of the Trapezium cluster, focusing our attention on 40 unresolved stellar sources with known spectral types and accurate WFPC2 UBVI-band photometry. We have estimated the ultraviolet excess considering two different reddening laws, and derived the stellar parameters and mass accretion rates using the D'Antona \& Mazzitelli (1998) and Palla \& Stahler 1999) computations. Approximately 75\% of the sources show appreciable excess luminosity in the U-band, that we attribute to accretion. We used the known correlation between the U-band excess and the total accretion luminosity (Gullbring et~al.\ 1998), recalibrated to our photometric system, to estimate the mass accretion rates, all found to be in the range $10^{-8}-10^{-12}\Msunyr$. We find some evidence for a decrease of $\dot{M}$ with the stellar age for sources older than 0.1~Myr, but in general the mass accretion rates appear to be lower than those measured in Taurus or in the flanking fields of the Orion Nebula. We suggest that the photoionization generated by the Trapezium OB stars causes a drop of the disk mass accretion rate. Low mass
stars therefore conclude their pre-main sequence evolution with lower masses, and the Initial Mass Function turns out to be affected by the rapid evolution of the most massive cluster's stars, with a surplus of ``accretion aborted'' stars or brown dwarfs, and a deficit of intermediate mass stars. This trend is in agreement with recent observations of the IMF in the Trapezium cluster.

\acknowledgments
The authors are indebted to Ed Fitzpatrick for discussions on the reddening law, to Francesco Palla and Keivan Stassun for their comments on an earlier version of the manuscript, to Charlie Lada and Michael Meyer for discussions, and to B.~Hilbert and M.~Richardson for their collaboration. J.~S.\ and G.~M.~C.\ have been supported by the Summer Student Program of the Space Telescope Science Institute. Support for J.~S.\  has been also provided by HST/DDRF grant 82316.

\begin{figure}
\label{figure1}
\epsscale{1}
\plotone{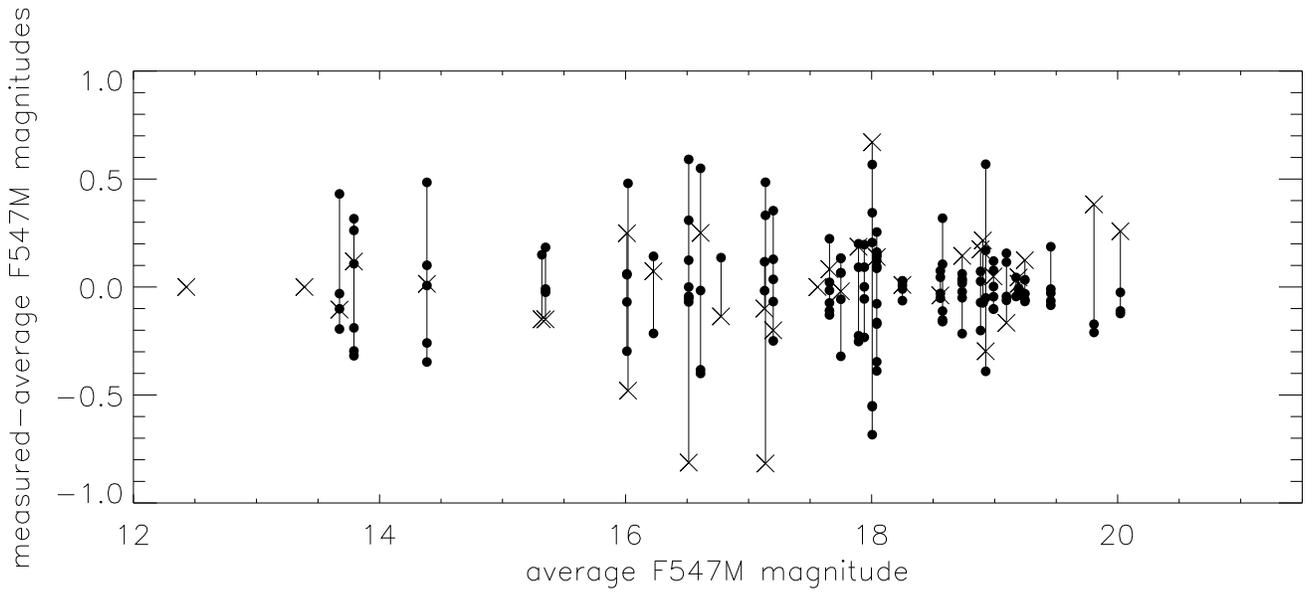}
\caption{Scatter of the measured $F547M$ (V-band) magnitudes vs.\ the mean values. Filled dots refer to WFPC2 data, crosses refer to the V-band data of Hillenbrand (1997).}
\end{figure}

\clearpage
\begin{figure}
\label{figure2}
\epsscale{0.85}
\plotone{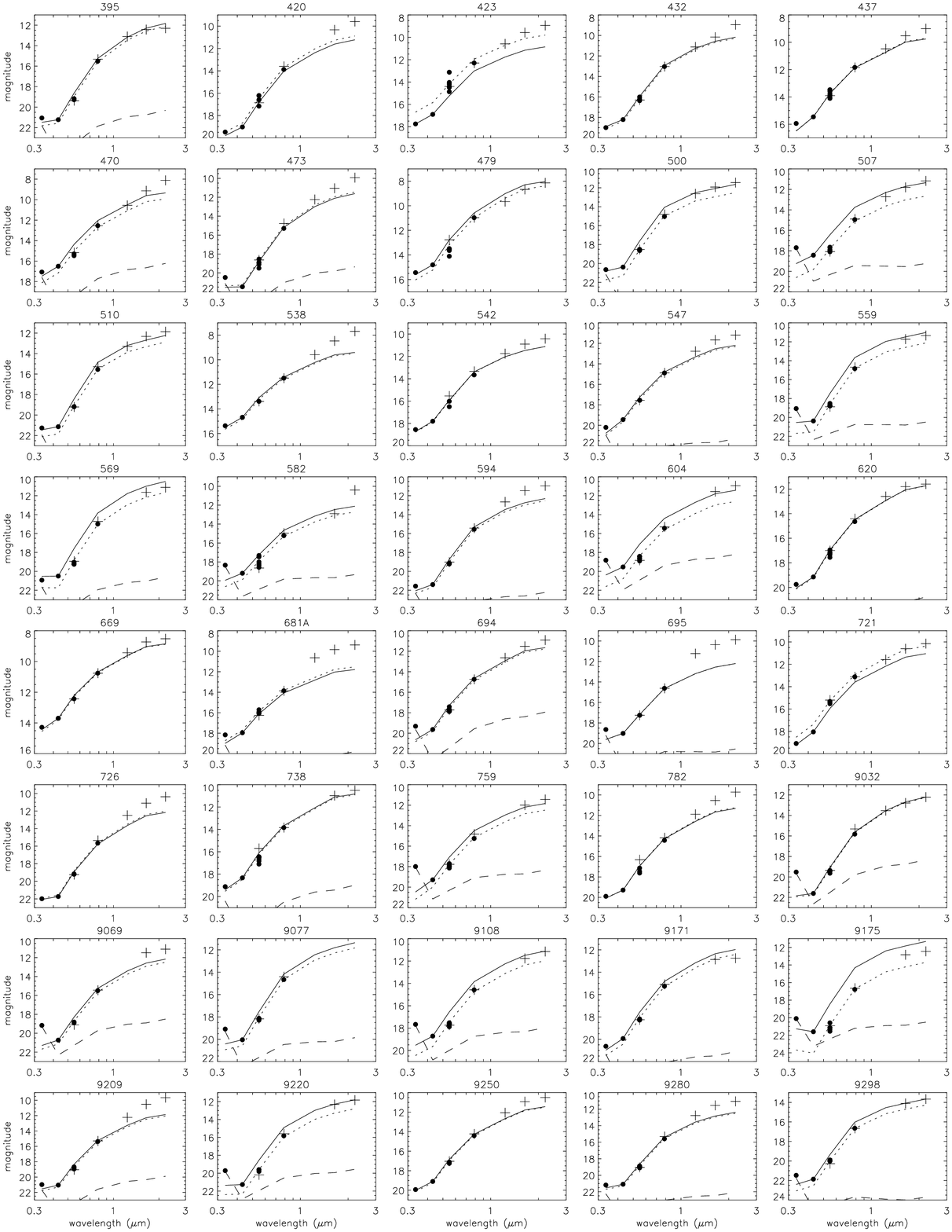}
\caption{Observed Spectral Energy Distributions and model results for the $R_V=3.1$ reddening law. The data are from our WFPC2 observations (points) and from the literature (plusses). We show the best fit for photosphere+accretion to the BVI data (dotted), this same fit translated to match the F439 data
(solid), and the accretion component alone (dashed).}

\end{figure}

\clearpage
\begin{figure}
\label{figure3}
\epsscale{0.85}
\plotone{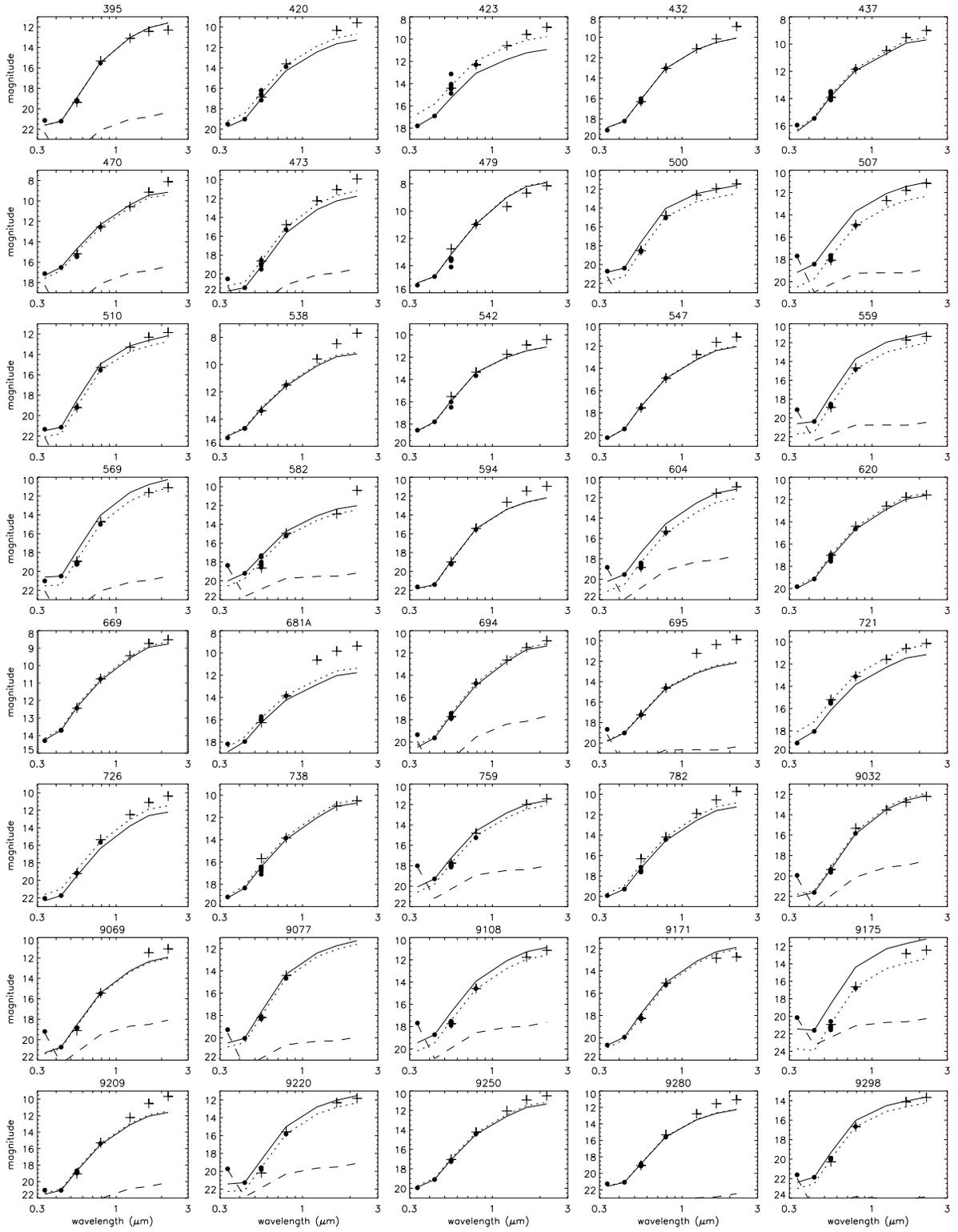}
\caption{Same as Figure~3 for $R_V=5.5$.}
\end{figure}

\clearpage
\begin{figure}
\label{figure4}
\epsscale{0.65}
\plotone{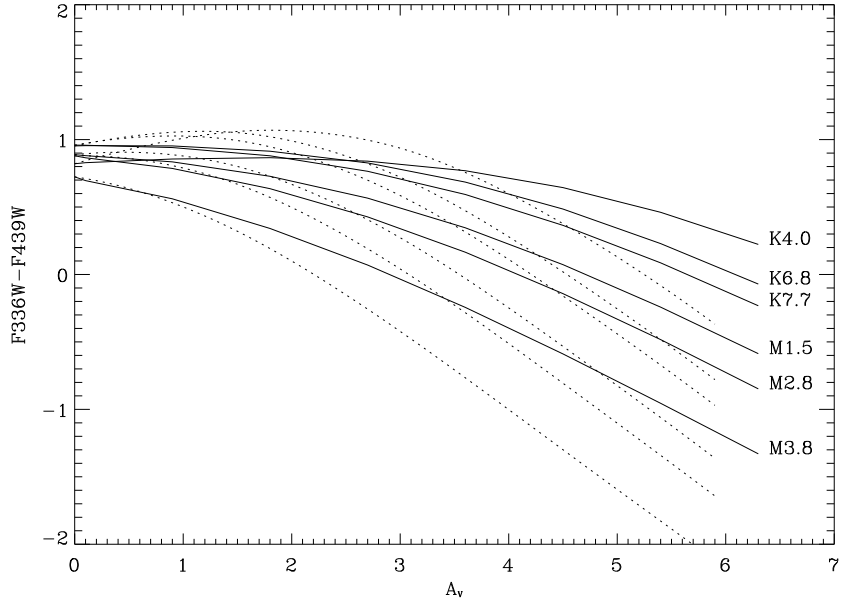}
\caption{Effect of the red-leak in the F336W filter, shown has a variation of the F336W-F439W color vs.\  extinction $A_V$ for our template stars. Color is bluer to the bottom. Solid lines are used for $R_V=5.5$, dotted lines for $R=3.1$. Labels at the right edge refer to the logarithm of the stellar effective temperature.}
\end{figure}

\begin{figure}
\label{figure5}
\epsscale{0.65}
\plotone{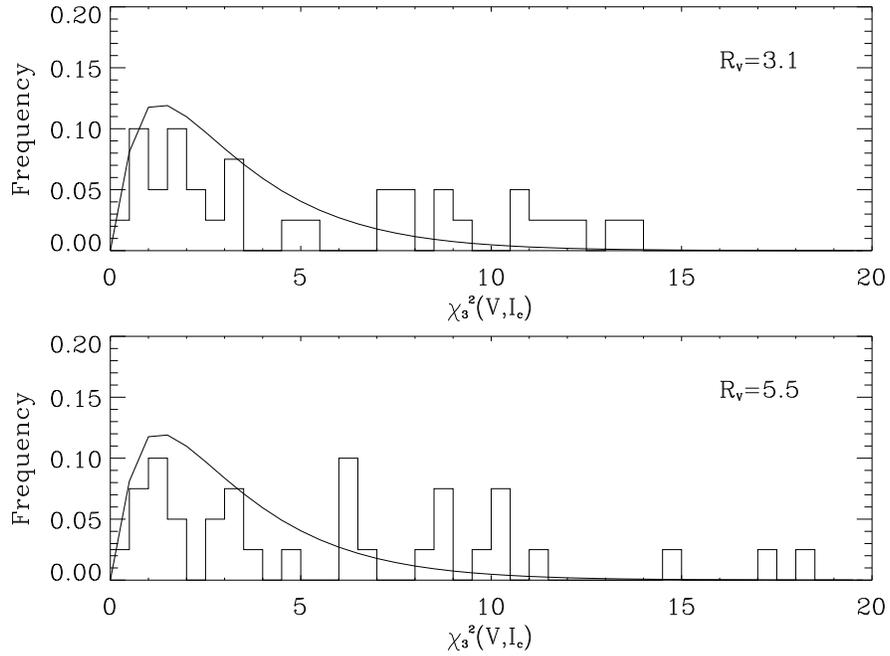}
\caption{Histogram of the minimum$\chi^2$ distribution, for both reddening laws. 
The solid lines refer to the theorical $\chi_2$ reduced to 3 degrees of freedom.}
\end{figure}

\clearpage
\begin{figure}
\epsscale{1.}
\label{figure6}
\plotone{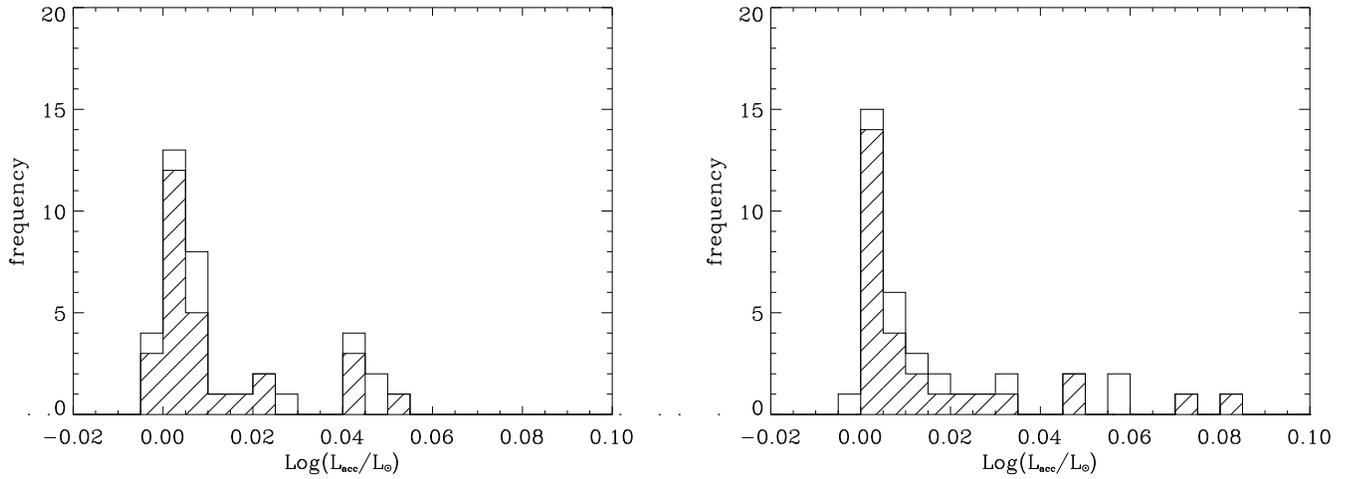}
\caption{Histogram of the F336W excess, after subtraction of the photospheric continuum. Bluer colors are to the right. Open areas are used for the entire sample of 40 stars, hatched areas for sources of the main sample.}
\end{figure}

\begin{figure}
\label{figure7}
\epsscale{.8}
\plotone{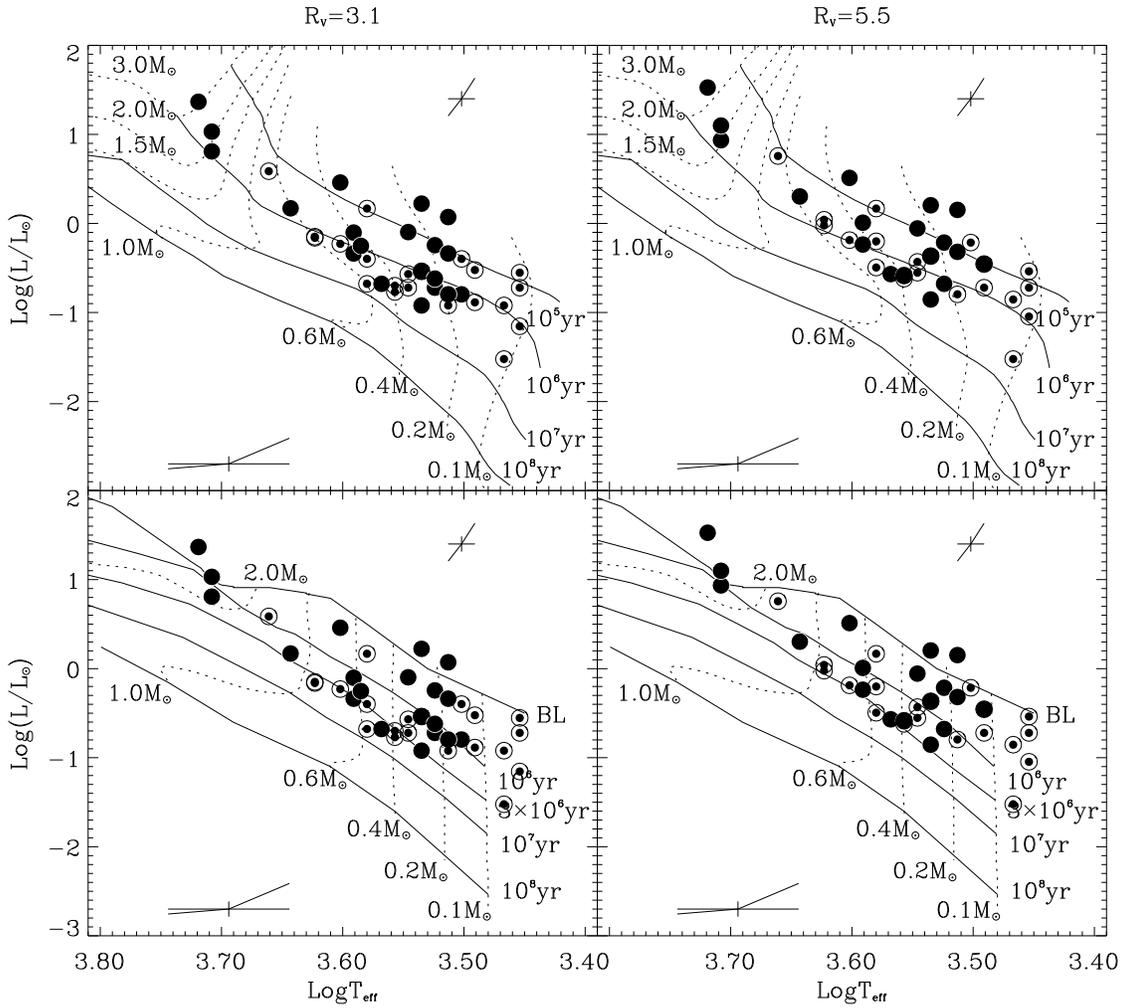}
\caption{HR diagrams obtained for different reddening laws ($R_V=3.1$, left colum; $R_V=5.5$, right column)
and evolutionary models (DM98, top row; PS99, bottom row). Filled circles are used for category~1 (most reliable fit) sources, open circles with inner dot are used for category~2 sources (less reliable fit). Ages and masses are indicated with conventional notation, with ``BL'' in the PS99 diagrams indicating the $10^{-5}\Msunyr$ birthline. The diagrams at $\log T_\mathrm{eff}=3.70, \log(L/L_\odot)=-2.8$ and $\log T_\mathrm{eff}=3.50, \log(L/L_\odot)=1.5$ represent the typical errors resulting from spectral classification, stellar variability, and bolometric correction (see text).}
\end{figure}

\clearpage
\begin{figure}
\label{figure8}
\epsscale{0.8}
\plotone{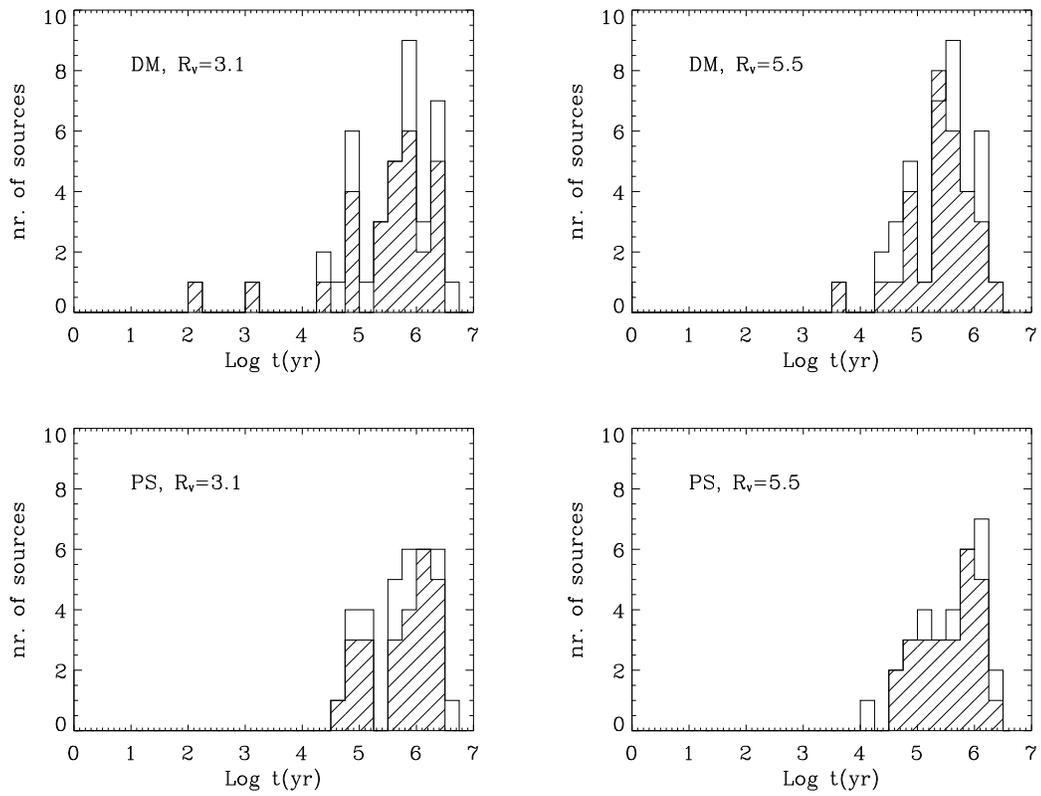}
\caption{Histogram of the age distribution. Open areas are used for the entire sample of 40 stars, hatched areas for sources of the main sample. Top row: DM98 model; bottom row: PS99 model; left column: $R_V=3.1$ reddening law; right column: $R_V=5.5$ reddening law.}
\end{figure}

\clearpage
\begin{figure}
\label{figure9}
\epsscale{1}
\plotone{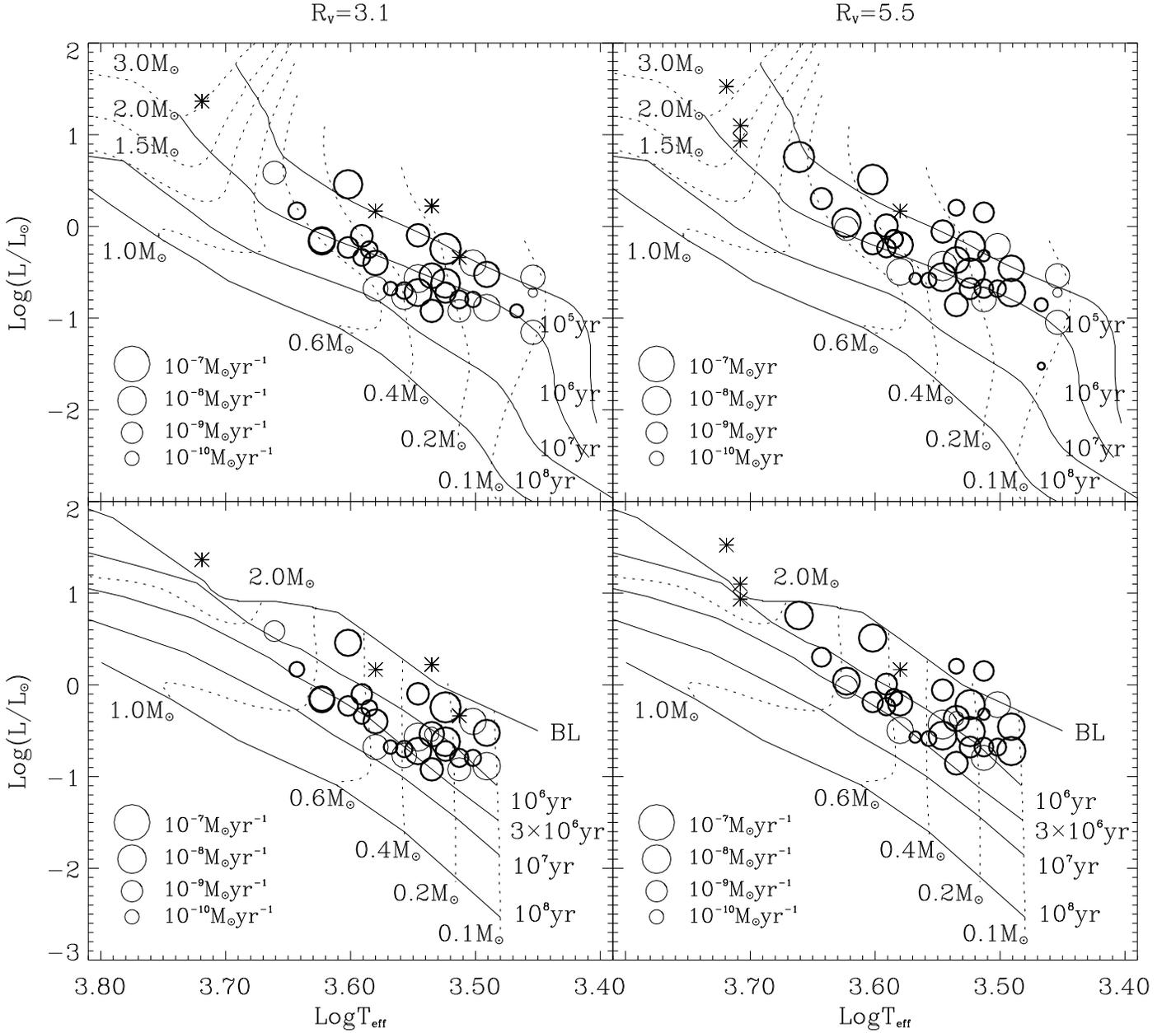}
\caption{HR diagrams with mass accretion rates. This figure, similar to Figure~7, represents  each star with an open circles with diameter proportional to the logarithm of the mass accretion rates. Asterisks 
are used for sources with UV deficit.}
\end{figure}

\clearpage
\begin{figure}
\label{figure10}
\epsscale{0.6}
\plotone{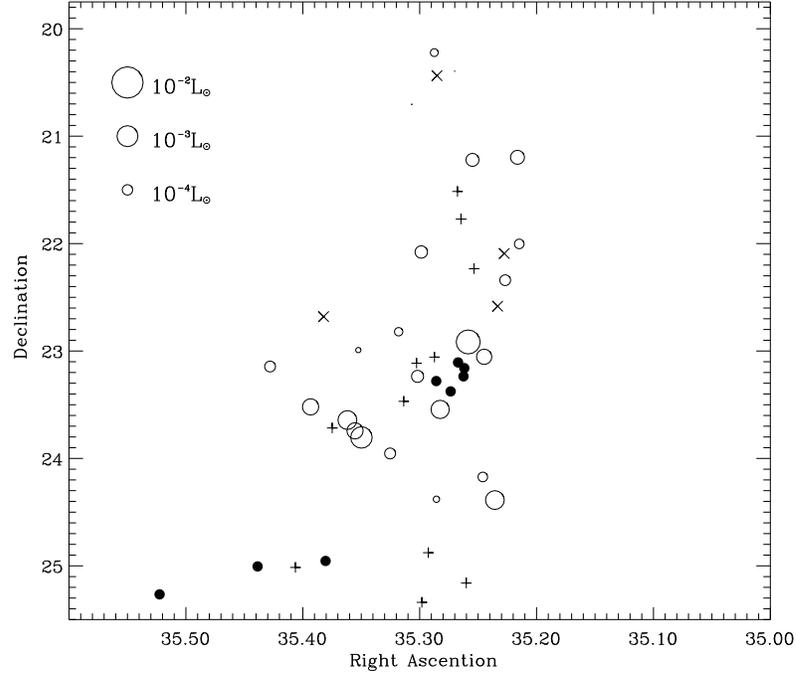}
\caption{Location of the accreting protostars in the Trapezium.Open circles, used for category~1 sources, are proportional to the logarithm of the F336W excess luminosity, estimated assuming $R_V=3.1$ reddening law; plus (+) signs are used for category~2 sources; times ({$\times$}) signs are used for sources with deficit of F336W emission. Filled dots represent cluster's OB stars, with the five Trapezium stars clearly visible at the center of the figure.}
\end{figure}

\begin{figure}
\epsscale{0.6}
\label{figure11}
\plotone{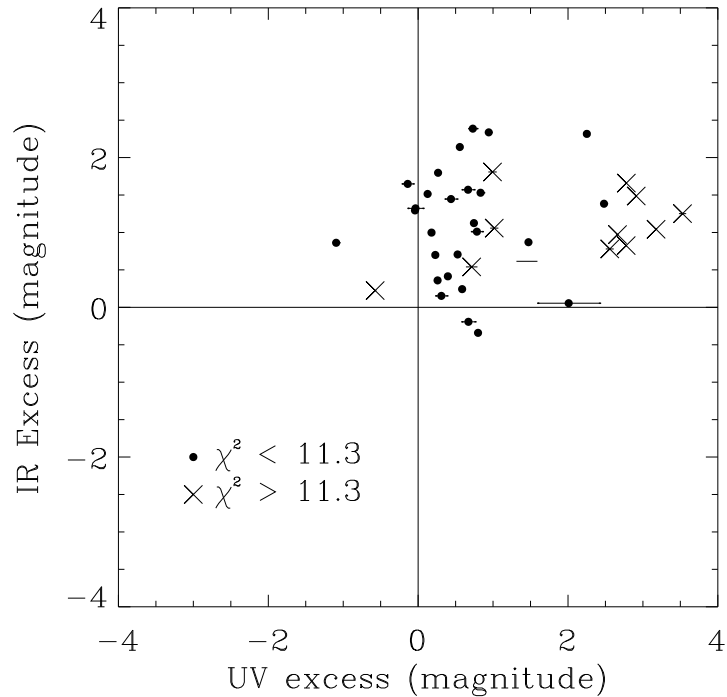}
\caption{Infrared (K-band) vs. UV excess, calculated assuming the $R_V=3.1$ reddening law. Filled circles indicate category~1 sources, times ($\times$) signs indicate category~2 sources. The error bars on the UV excess represent the uncertainty associated to the U-band photometric errors.}
\end{figure}

\clearpage
\begin{figure}
\epsscale{.8}
\label{figure12}
\plotone{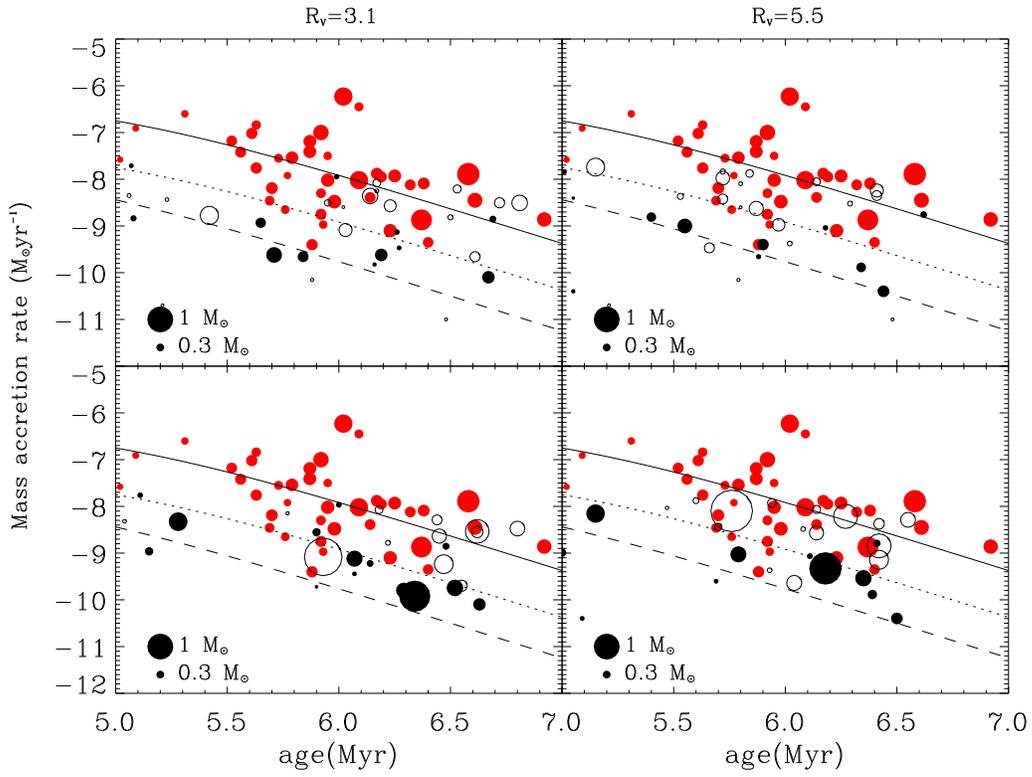}
\caption{Mass accretion rate vs.\ age for the main sample. The four diagrams represent different choices for the reddening law ($R_V=3.1$, left column; $R_V=5.5$, right column) and for the evolutionary models (DM98, top row; PS99, bottom row).  {\sl Dark filled circles}: main sources; {\sl open circles}: secondary sources; {\sl red filled circles} mass accretion rates estimated by Hartmann et~al.\ (1998) for the Taurus cluster. {\sl Solid line}: similarity solution from the fiducial model of Hartmann et~al.\ (1998); {\sl Dotted line}: similarity solution, with $M_\star=0.1$~\Msun; {\sl Dashed line}: similarity solution, with $M_\star=0.1$~\Msun\, and $M_d(0)=0.01$~\Msun.}
\end{figure}

\clearpage
\begin{deluxetable}{ccccccccc}
\label{table1}
\tabletypesize{\scriptsize}
\tablewidth{0pt}
\tablecolumns{9}
\tablecaption{WFPC2 photometry}
\tablehead{\colhead{Star ID\tablenotemark{a}} & \colhead{R.A.(2000.0)} & \colhead{Dec(2000.0)} &
   \colhead{F336W} & \colhead{F439W} &  \colhead{F547M} & \colhead{F791W} &
   \colhead{Nr. Obs.} & \colhead{$\sigma$(F547M)}}
\startdata
395  & $5^h 35^m 12\fs89$ & $-5\degr22\arcmin0\farcs19$ & $21.144\pm0.087 $&$ 21.225\pm0.044 $&$ 19.215\pm0.018 $&$ 15.704\pm0.001$ & 5 & 0.080\\
420 & $5^h 35^m 13\fs61$ & $-5\degr22\arcmin20\farcs45$ & $19.500\pm0.019 $&$ 19.024\pm0.031 $& $16.343\pm0.002$ &$14.162\pm0.013$ & 5 & 0.410\\
423 & $5^h 35^m 13\fs66$ & $-5\degr22\arcmin05\farcs51$ & $17.789\pm0.037 $&$ 16.899\pm0.033 $&$14.261\pm0.003$& \nodata & 6 & 0.294\\
432 & $5^h 35^m 14\fs00$ & $-5\degr22\arcmin34\farcs99$ & $19.099\pm0.100 $&$ 18.213\pm0.057 $& $16.191\pm0.002$ & \nodata & 3 & 0.189\\
437 & $5^h 35^m 14\fs14$ & $-5\degr24\arcmin23\farcs24$ & $15.952\pm0.008 $&$ 15.462\pm0.002 $&$13.772\pm0.003$ & \nodata & 7 & 0.264 \\
470 & $5^h 35^m 15\fs21$ & $-5\degr22\arcmin14\farcs04$ & $17.122\pm0.056 $&$ 16.503\pm0.085 $&$ 14.427\pm0.001$&\nodata & 2 & 0.211 \\
473  & $5^h 35^m 15\fs29$ & $-5\degr21\arcmin13\farcs25$ &$20.523\pm0.041 $&$ 21.463\pm0.053$&$ 19.002\pm0.009 $&$ 15.561\pm0.001$ & 5 & 0.386\\
479 & $5^h 35^m 15\fs51$ & $-5\degr22\arcmin54\farcs94$& $ 15.439\pm0.002 $ & $14.805\pm0.001$ &$13.609\pm0.003$ & \nodata & 5 & 0.248\\
500  & $5^h 35^m 15\fs88$ & $-5\degr21\arcmin46\farcs31$ &    $20.706\pm0.055 $&$ 20.382\pm0.022 $&$ 18.569\pm0.008 $&$ 15.146\pm0.002$ & 5 & 0.056\\
507 & $5^h 35^m 16\fs06$ & $-5\degr21\arcmin30\farcs88$ & $17.695\pm0.005 $&$ 18.427\pm0.006 $& $17.847\pm0.006$ &$15.004\pm0.002$ & 5 & 0.222\\
510 & $5^h 35^m 16\fs20$ & $-5\degr20\arcmin23\farcs70$ & $21.333\pm0.072 $&$ 21.137\pm0.033 $&$19.181\pm0.051$&$15.817\pm0.004$ & 2 & 0.021\\
538 & $5^h 35^m 16\fs95$ & $-5\degr23\arcmin32\farcs64$ &$15.383\pm0.010 $&$ 14.687\pm0.002 $& \nodata & \nodata & 1 & \nodata\\
542 & $5^h 35^m 17\fs11$ & $-5\degr20\arcmin26\farcs20$ & $18.559\pm0.012 $&$ 17.804\pm0.004 $&$16.500\pm0.003$&$13.980\pm0.004$ & 2 & 0.679\\
547 & $5^h 35^m 17\fs25$ & $-5\degr20\arcmin13\farcs33$ & $20.230\pm0.023 $&$ 19.446\pm0.010 $& \nodata & \nodata & 1 & \nodata\\
559  & $5^h 35^m 17\fs56$ & $-5\degr24\arcmin52\farcs66$ & $19.120\pm0.052 $&$ 20.377\pm0.153 $&$18.910\pm0.012$&$14.991\pm0.001$ & 8 & 0.105\\
569 & $5^h 35^m 17\fs89$ & $-5\degr25\arcmin20\farcs38$ & $21.007\pm0.077 $&$ 20.497\pm0.029 $&$19.137\pm0.013$&$15.259\pm0.002$ & 5 & 0.133\\
582 & $5^h 35^m 18\fs11$ & $-5\degr23\arcmin14\farcs17$ & $18.362\pm0.017 $&$ 19.200\pm0.017 $&$17.893\pm0.006$&$15.492\pm0.001$ & 7 & 0.578\\
594  & $5^h 35^m 18\fs41$ & $-5\degr20\arcmin42\farcs25$ &    $21.639\pm0.079 $&$ 21.394\pm0.039 $&$ 19.174\pm0.010 $&$ 15.671\pm0.001$ & 2 & 0.062\\
604 & $5^h 35^m 18\fs82$ & $-5\degr23\arcmin28\farcs12$ & $18.830\pm0.010 $&$ 19.532\pm0.012 $&$ 18.576\pm0.038 $&$ 15.572\pm0.001$ & 5 & 0.208\\
620  & $5^h 35^m 19\fs52$ & $-5\degr23\arcmin57\farcs23$ & 	$18.828\pm0.074 $&$ 19.145\pm0.090 $&$ 17.241\pm0.005 $&$ 14.877\pm0.001$ & 6 & 0.223\\
669 & $5^h 35^m 20\fs99$ & $-5\degr23\arcmin48\farcs27$&$ 14.285\pm0.004 $&$ 13.697\pm0.001 $& \nodata & \nodata & 1 & \nodata\\
681& $5^h 35^m 21\fs32$ & $-5\degr23\arcmin44\farcs58$ & $18.169\pm0.007 $&$ 17.957\pm0.071 $& $15.949\pm0.004$& \nodata &  5 & 0.201\\
694 & $5^h 35^m 21\fs71$ & $-5\degr23\arcmin38\farcs58$ &$19.344\pm0.025 $&$ 19.638\pm0.015$&$17.753\pm0.008$ & \nodata & 7 & 0.158\\
695 & $5^h 35^m 21\fs67$ & $-5\degr26\arcmin43\farcs63$ & $18.49\pm0.03 $&$ 18.87\pm0.02 $& \nodata & \nodata & 1 & \nodata\\
721 & $5^h 35^m 22\fs49$ & $-5\degr23\arcmin42\farcs98$ &$19.099\pm0.014 $&$ 18.048\pm0.005 $& $15.399\pm0.004$ & \nodata & 4 & 0.137\\
726 & $5^h 35^m 22\fs93$ & $-5\degr22\arcmin40\farcs80$ &$22.059\pm0.0073 $&$ 21.734\pm0.041 $&$19.174\pm0.021$ & $15.960\pm0.001$ & 3 & 0.042\\
738 & $5^h 35^m 23\fs60$ & $-5\degr23\arcmin31\farcs21$ & $19.134\pm0.014 $&$ 18.333\pm0.010 $& $16.543\pm0.005$ &$14.583\pm0.003$ & 9 & 0.376\\
759  & $5^h 35^m 24\fs38$ & $-5\degr25\arcmin0\farcs88$ &      $18.003\pm0.015 $&$ 19.283\pm0.020 $&$ 17.940\pm0.011 $&$ 15.683\pm0.006$ & 5 & 0.161\\
782 & $5^h 35^m 25\fs68$ & $-5\degr23\arcmin08\farcs66$ & $19.920\pm0.026 $&$ 19.286\pm0.018 $& $17.546\pm0.012$ &$14.686\pm0.001$ & 3 & 0.711\\
9032  & $5^h 35^m 12\fs98$ & $-5\degr21\arcmin11\farcs82$ &   $19.938\pm0.412 $&$ 21.614\pm0.060$&$ 19.456\pm0.015 $&$ 15.997\pm0.001$ & 5 & 0.108\\
9069& $5^h 35^m 14\fs69$ & $-5\degr23\arcmin03\farcs23$ & $19.200\pm0.032 $&$ 20.728\pm0.048 $&$18.833\pm0.012$&$15.541\pm0.015$ & 4 & 0.143\\
9077& $5^h 35^m 14\fs76$ & $-5\degr24\arcmin10\farcs35$ & $19.27\pm0.18 $&$ 20.047\pm0.020 $&$18.186\pm0.006$&$14.928\pm0.003$ & 13 & 0.207\\
9108& $5^h 35^m 15\fs61$ & $-5\degr25\arcmin09\farcs61$ & $17.677\pm0.005 $&$ 18.721\pm0.010 $& $17.644\pm0.005$&\nodata & 7 & 0.124\\
9171& $5^h 35^m 17\fs14$ & $-5\degr24\arcmin22\farcs79$ & $20.65\pm0.02 $&$ 19.938\pm0.019 $&$18.248\pm0.008$ & $15.421\pm0.002$ & 6 & 0.033\\
9175& $5^h 35^m 17\fs24$ & $-5\degr23\arcmin03\farcs41$ & $20.148\pm0.048 $&$ 21.601\pm0.071 $&$21.149\pm0.352$&$16.912\pm0.004$ & 2 & 0.192\\
9209 & $5^h 35^m 17\fs92$ & $-5\degr22\arcmin 4\farcs70$ & $21.041\pm0.054 $&$ 21.057\pm0.032 $&$ 18.895\pm0.017 $&$ 15.446\pm0.003$ & 5 & 0.143\\
9220 & $5^h 35^m 18\fs16$ & $-5\degr23\arcmin6\farcs80$ &     $19.730\pm0.012 $&$ 21.270\pm0.052 $&$ 19.616\pm0.038 $&$ 15.940\pm0.011$ & 3 & 0.332\\
9250& $5^h 35^m 19\fs08$ & $-5\degr22\arcmin49\farcs23$ & $19.947\pm0.020 $&$ 19.099\pm0.008 $& $17.180\pm0.003$ & \nodata & 3 & 0.109\\    
9280 & $5^h 35^m 21\fs15$ & $-5\degr22\arcmin59\farcs49$ &      $21.262\pm0.078 $&$ 21.089\pm0.031$&$ 18.983\pm0.012$&$ 15.716\pm0.001$ & 7 & 0.087\\
9298 & $5^h 35^m 22\fs04$ & $-5\degr22\arcmin12\farcs77$ &       $21.610\pm0.140 $&$ 21.862\pm0.051 $&$ 19.936\pm0.020 $&$ 16.642\pm0.002$ & 4& 0.178\\
\enddata
\tablenotetext{a}{from Hillenbrand 1997}
\end{deluxetable}


\clearpage
\begin{deluxetable}{cccccccc}
\label{table2}
\tablewidth{0pt}
\tablecaption{Template stars from the Bruzual-Persson-Gunn-Stryker Spectral Atlas}
\tablehead{\colhead{BPGS} & \colhead{Name} & \colhead{V-I$_c$} & \colhead{Sp.T.\tablenotemark{a}} & \colhead{$\log $T$_{eff}$\tablenotemark{a}}}
\startdata
48& HD~139777	&	0.758	& G6.2V	&	3.741	\\
56& HD~190470 	& 	0.984	& K1.6V 	&	3.697	\\
57& HD~154712 	& 	1.130	& K3.3V 	&	3.672	\\
59& BD+38~2457	&	1.206	& K4.0V	&	3.660	\\
61& Gl~40		&	1.565	& K6.8V	&	3.605	\\
65& HD~132683	& 	1.693	& K7.7V 	&	3.594	\\
67& GL~49 		& 	2.147	& M1.5V 	&	3.549	\\
68& GL~109 		& 	2.474	& M2.8V 	&	3.530	\\
69& GL~15B 		& 	2.835	& M3.8V 	&	3.512	\\
\enddata
\tablenotetext{a}{spectral types and effective temperatures derived from the $V-I_c$ color according to Hillenbrand (1997)  (see text).}
\end{deluxetable}

\begin{deluxetable}{ccccccccccccc}
\label{table3}
\tablecolumns{12}
\tablecaption{Stellar and Mass Accretion Parameters for $R_V=3.1$}
\tablehead{
\colhead{} & \colhead{} & \colhead{} & \colhead{} & \colhead{} & \colhead{} & 
\multicolumn{3}{c}{Palla \& Stahler (1999)} & \colhead{} & 
\multicolumn{3}{c}{D'Antona \& Mazzitelli (1998)} \\
\cline{7-9}  \cline{11-13}\\
\colhead{ID}
	& \colhead{T$_{eff}$}
		& \colhead{L$_\star$} 
			& \colhead{$A_V$} 
				& \colhead{R$_\star$} 
						& \colhead{L$_{accretion}$} 
							& \colhead{M$_\star$} 
								& \colhead{age} 
									& \colhead{$\dot{M}$}  
						& \colhead{} 
							& \colhead{M$_\star$} 
								& \colhead{age} 
									& \colhead{$\dot{M}$}  \\
\colhead{} 
	& \colhead{(\scriptsize K)} 
			& \colhead{\scriptsize(L$_\odot$)} 
				&  \colhead{(mag)} 
					& \colhead{\scriptsize(R$_\odot$)} 
						& \colhead{\scriptsize(L$_\odot$)} 
							& \colhead{\scriptsize(M$_\odot$)} 
								& \colhead{(yr)} 
									& \colhead{\scriptsize($10^{-9}$M$_\odot$ yr$^{-1}$)} 
					& \colhead{} 
							& \colhead{\scriptsize(M$_\odot$)} 
								& \colhead{(yr)} 
									& \colhead{\scriptsize($10^{-9}$M$_\odot$ yr$^{-1}$)} }
\startdata
395	&	3427.7	&	0.29	&	2.96	&	1.54	&	1.27E-02	&	0.28	&	5.90	&	2.79	&	&	0.24	&	5.95	&	3.12	\\
420	&	3515.6	&	0.80	&	1.71	&	2.42	&	3.68E-03	&	0.32	&	5.15	&	1.09	&	&	0.24	&	5.08	&	1.46	\\
423	&	3427.7	&	1.67	&	0.14	&	3.68	&	-2.33E-04	&	0.26	&\nodata	&	\nodata	&	&	0.18	&	2.46	&	\nodata	\\
432	&	3258.4	&	1.18	&	1.30	&	3.42	&	-1.48E-04	&	0.17	&\nodata	&	\nodata	&	&	0.15	&	3.33	&	\nodata	\\
437	&	3999.5	&	2.88	&	0.79	&	3.55	&	2.41E-02	&	0.72	&	5.28	&	4.70	&	&	0.34	&	4.67	&	9.78	\\
470	&	4581.4	&	3.86	&	2.85	&	3.13	&	9.63E-03	&	1.45	&	5.93	&	0.83	&	&	0.70	&	5.42	&	1.68	\\
473	&	3342.0	&	0.57	&	3.18	&	2.26	&	4.23E-02	&	0.22	&	5.11	&	17.3	&	&	0.19	&	5.07	&	19.5	\\
479	&	5236.0	&	23.25	&	3.30	&	5.88	&	-5.14E+00	&	2.97	&	5.47	&	\nodata	&	&	3.25	&	6.64	&	\nodata	\\
500	&	2844.5	&	0.19	&	0.09	&	1.80	&	3.07E-05	&\nodata	&\nodata	&	\nodata	&	&	0.11	&	5.21	&	0.02	\\
507	&	3258.4	&	0.12	&	0.50	&	1.08	&	7.18E-03	&	0.19	&	6.22	&	1.66	&	&	0.20	&	6.50	&	1.53	\\
510	&	2930.9	&	0.12	&	0.69	&	1.37	&	1.53E-04	&\nodata	&\nodata	&	\nodata	&	&	0.12	&	5.88	&	0.07	\\
538	&	5105.1	&	6.44	&	1.91	&	3.26	&	-4.67E-01	&	2.04	&	5.94	&	\nodata	&	&	1.92	&	5.94	&	\nodata	\\
542	&	3258.4	&	0.46	&	0.16	&	2.13	&	-2.13E-04	&	0.18	&	5.17	&	\nodata	&	&	0.17	&	5.10	&	\nodata	\\
547	&	3698.3	&	0.21	&	1.41	&	1.12	&	8.65E-04	&	0.48	&	6.63	&	0.08	&	&	0.49	&	6.67	&	0.08	\\
559	&	2844.5	&	0.28	&	0.62	&	2.17	&	3.18E-03	&\nodata	&\nodata	&	\nodata	&	&	0.10	&	4.56	&	2.61	\\
569	&	3176.9	&	0.40	&	2.58	&	2.10	&	8.11E-03	&	0.14	&	5.04	&	4.80	&	&	0.15	&	5.06	&	4.47	\\
582	&	3427.7	&	0.12	&	1.00	&	1.00	&	9.58E-03	&	0.27	&	6.48	&	1.40	&	&	0.27	&	6.69	&	1.40	\\
594	&	3176.9	&	0.16	&	1.80	&	1.32	&	5.06E-04	&	0.14	&	5.90	&	0.19	&	&	0.17	&	6.16	&	0.15	\\
604	&	3801.9	&	0.21	&	2.31	&	1.05	&	4.55E-02	&	0.57	&	6.80	&	3.35	&	&	0.60	&	6.81	&	3.11	\\
620	&	3899.4	&	0.46	&	1.96	&	1.49	&	1.98E-03	&	0.64	&	6.52	&	0.18	&	&	0.49	&	6.19	&	0.24	\\
669	&	5105.1	&	10.75	&	1.54	&	4.21	&	-5.94E-01	&	2.23	&	5.35	&	\nodata	&	&	2.06	&	5.73	&	\nodata	\\
681	&	3999.5	&	0.59	&	1.18	&	1.60	&	6.58E-03	&	0.73	&	6.47	&	0.58	&	&	0.51	&	6.03	&	0.82	\\
694	&	4197.6	&	0.69	&	3.12	&	1.58	&	5.23E-02	&	0.92	&	6.64	&	3.55	&	&	0.60	&	6.16	&	5.36	\\
695	&	3342.0	&	0.19	&	0.61	&	1.31	&	3.10E-03	&	0.27	&	6.14	&	0.60	&	&	0.21	&	6.26	&	0.74	\\
721	&	3801.9	&	1.47	&	1.39	&	2.80	&	-2.56E-04	&	0.53	&	5.34	&	\nodata	&	&	0.30	&	4.96	&	\nodata	\\
726	&	4197.6	&	0.71	&	4.44	&	1.60	&	4.18E-02	&	0.91	&	6.62	&	2.92	&	&	0.60	&	6.14	&	4.39	\\
738	&	4395.4	&	1.48	&	2.90	&	2.10	&	1.78E-03	&	1.20	&	6.34	&	0.12	&	&	0.61	&	5.71	&	0.24	\\
759	&	3605.8	&	0.17	&	1.47	&	1.07	&	2.95E-02	&	0.44	&	6.62	&	2.83	&	&	0.39	&	6.72	&	3.14	\\
782	&	3899.4	&	0.79	&	2.43	&	1.96	&	6.12E-03	&	0.62	&	6.07	&	0.76	&	&	0.40	&	5.65	&	1.17	\\
9032	&	3342.0	&	0.24	&	2.81	&	1.47	&	4.10E-02	&	0.22	&	6.00	&	10.8	&	&	0.21	&	5.99	&	11.3	\\
9069	&	3515.6	&	0.19	&	2.36	&	1.18	&	4.18E-02	&	0.38	&	6.44	&	5.14	&	&	0.31	&	6.53	&	6.19	\\
9077	&	3097.4	&	0.30	&	1.26	&	1.90	&	6.85E-03	&	0.11	&	4.93	&	4.67	&	&	0.14	&	5.23	&	3.66	\\
9108	&	3515.6	&	0.27	&	1.55	&	1.41	&	4.56E-02	&	0.30	&	6.18	&	8.48	&	&	0.30	&	6.17	&	8.32	\\
9171	&	3605.8	&	0.20	&	1.91	&	1.14	&	1.92E-03	&	0.43	&	6.55	&	0.20	&	&	0.39	&	6.61	&	0.22	\\
9175	&	2844.5	&	0.07	&	1.41	&	1.11	&	6.12E-03	&\nodata	&\nodata	&	\nodata	&	&	0.11	&	6.02	&	2.53	\\
9209	&	3801.9	&	0.40	&	3.23	&	1.45	&	2.19E-02	&	0.55	&	6.45	&	2.31	&	&	0.46	&	6.23	&	2.71	\\
9220	&	3097.4	&	0.13	&	1.92	&	1.28	&	1.71E-02	&	0.12	&	5.77	&	7.10	&	&	0.15	&	6.17	&	5.61	\\
9250	&	3845.9	&	0.56	&	2.04	&	1.69	&	1.39E-03	&	0.59	&	6.29	&	0.16	&	&	0.42	&	5.84	&	0.22	\\
9280	&	3258.4	&	0.16	&	1.86	&	1.28	&	1.28E-03	&	0.18	&	6.07	&	0.36	&	&	0.19	&	6.27	&	0.34	\\
9298	&	2930.9	&	0.03	&	0.09	&	0.70	&	4.77E-05	&\nodata	&\nodata	&	\nodata	&	&	0.11	&	6.48	&	0.01	\\
\enddata
\end{deluxetable}

\clearpage
\begin{deluxetable}{ccccccccccccc}
\label{table4}
\tablecolumns{12}
\tablecaption{Stellar and Mass Accretion Parameters for $R_V=5.5$}
\tablehead{
\colhead{} & \colhead{} & \colhead{} & \colhead{} & \colhead{} & \colhead{} & 
\multicolumn{3}{c}{Palla \& Stahler (1999)} & \colhead{} & 
\multicolumn{3}{c}{D'Antona \& Mazzitelli (1998)} \\
\cline{7-9}  \cline{11-13}\\
\colhead{ID}
	& \colhead{T$_{eff}$}
		& \colhead{L$_\star$} 
			& \colhead{$A_V$} 
				& \colhead{R$_\star$} 
						& \colhead{L$_{accretion}$} 
							& \colhead{M$_\star$} 
								& \colhead{age} 
									& \colhead{$\dot{M}$}  
						& \colhead{} 
							& \colhead{M$_\star$} 
								& \colhead{age} 
									& \colhead{$\dot{M}$}  \\
\colhead{} 
	& \colhead{(\scriptsize K)} 
			& \colhead{\scriptsize(L$_\odot$)} 
				&  \colhead{(mag)} 
					& \colhead{\scriptsize(R$_\odot$)} 
						& \colhead{\scriptsize(L$_\odot$)} 
							& \colhead{\scriptsize(M$_\odot$)} 
								& \colhead{(yr)} 
									& \colhead{\scriptsize($10^{-9}$M$_\odot$ yr$^{-1}$)} 
					& \colhead{} 
							& \colhead{\scriptsize(M$_\odot$)} 
								& \colhead{(yr)} 
									& \colhead{\scriptsize($10^{-9}$M$_\odot$ yr$^{-1}$)} }
\startdata
395	&	3427.7	&	0.43	&	3.36	&	1.86	&	1.39E-02	&	0.28	&	5.70	&	3.64	&	&	0.24	&	5.53	&	4.30	\\
420	&	3515.6	&	1.75	&	2.66	&	3.58	&	8.26E-04	&	0.33	&	4.51	&	0.36	&	&	0.20	&	6.17	&	0.59	\\
423	&	3427.7	&	3.92	&	1.40	&	5.63	&	-2.43E-03	&	0.27	&\nodata	&	\nodata	&	&	0.13	&	\nodata	&	\nodata	\\
432	&	3258.4	&	1.12	&	0.55	&	3.33	&	-4.44E-04	&	0.17	&\nodata	&	\nodata	&	&	0.15	&	3.42	&	\nodata	\\
437	&	3999.5	&	3.36	&	1.06	&	3.83	&	3.83E-02	&	0.72	&	5.13	&	8.08	&	&	0.34	&	4.54	&	17	\\
470	&	4581.4	&	3.59	&	2.61	&	3.02	&	3.92E-02	&	1.49	&	5.99	&	3.16	&	&	0.67	&	5.48	&	6.88	\\
473	&	3342.0	&	0.96	&	3.65	&	2.94	&	6.63E-03	&	0.21	&	4.32	&	3.69	&	&	0.18	&	5.19	&	4.34	\\
479	&	5236.0	&	33.35	&	3.69	&	7.05	&	-9.64E-01	&	2.68	&\nodata	&	\nodata	&	&	4.28	&	6.18	&	\nodata	\\
500	&	2844.5	&	3.45	&	0.01	&	7.68	&	-1.91E-05	&\nodata	&\nodata	&	\nodata	&	&	\nodata	&	\nodata	&	\nodata	\\
507	&	3258.4	&	0.47	&	0.01	&	2.15	&	3.92E-03	&	0.19	&	5.14	&	1.80	&	&	0.17	&	5.08	&	1.97	\\
510	&	2930.9	&	0.62	&	0.01	&	3.08	&	1.10E-05	&\nodata	&\nodata	&	\nodata	&	&	0.11	&	\nodata	&0.01	\\
538	&	5105.1	&	8.3	&	2.24	&	3.70	&	-1.11E-01	&	2.04	&	5.54	&	\nodata	&	&	2.00	&	5.84	&	\nodata	\\
542	&	3258.4	&	0.9	&	0.01	&	2.98	&	-2.92E-05	&	0.16	&	4.21	&	\nodata	&	&	0.16	&	5.11	&	\nodata	\\
547	&	3698.3	&	0.32	&	1.74	&	1.37	&	2.72E-05	&	0.47	&	6.37	&	0.003	&	&	0.43	&	6.28	&	0.003	\\
559	&	2844.5	&	3.04	&	0.01	&	7.21	&	8.72E-04	&\nodata	&\nodata	&	\nodata	&	&	\nodata	&	\nodata	&	\nodata	\\
569	&	3176.9	&	0.17	&	0.28	&	1.37	&	1.32E-05	&	0.15	&	5.85	&	0.005	&	&	0.17	&	6.10	&	0.004	\\
582	&	3427.7	&	0.09	&	0.01	&	0.86	&	2.10E-03	&	0.36	&	6.69	&	0.20	&	&	0.28	&	6.77	&	0.25	\\
594	&	3176.9	&	0.19	&	1.13	&	1.44	&	2.57E-05	&	0.17	&	5.77	&	0.01	&	&	0.16	&	5.99	&	0.01	\\
604	&	3801.9	&	0.07	&	0.74	&	0.63	&	3.17E-03	&	0.47	&	7.00	&	0.17	&	&	0.58	&	7.48	&	0.13	\\
620	&	3899.4	&	0.66	&	2.48	&	1.79	&	6.92E-04	&	0.64	&	6.26	&	0.08	&	&	0.43	&	5.82	&	0.11	\\
669	&	5105.1	&	13.12	&	1.83	&	4.65	&	-4.23E-02	&	2.54	&	5.40	&	\nodata	&	&	2.02	&	5.64	&	\nodata	\\
681	&	3999.5	&	0.88	&	1.93	&	1.96	&	1.50E-02	&	0.72	&	6.22	&	1.62	&	&	0.44	&	5.69	&	2.61	\\
694	&	4197.6	&	1.08	&	3.81	&	1.98	&	6.53E-02	&	0.93	&	6.28	&	5.52	&	&	0.53	&	5.74	&	9.53	\\
695	&	3342.0	&	0.17	&	0.01	&	1.23	&	1.22E-03	&	0.25	&	6.20	&	0.240	&	&	0.22	&	6.38	&	0.27	\\
721	&	3801.9	&	4.23	&	3.09	&	4.76	&	-1.90E-02	&	0.53	&\nodata	&	\nodata	&	&	0.25	&	5.09	&	\nodata	\\
726	&	4197.6	&	2.29	&	6.14	&	2.87	&	-7.11E-03	&	0.94	&	5.82	&	\nodata	&	&	0.45	&	5.18	&	\nodata	\\
738	&	4395.4	&	2.55	&	3.62	&	2.77	&	2.90E-03	&	1.29	&	6.00	&	0.25	&	&	0.55	&	5.43	&	0.57	\\
759	&	3605.8	&	0.14	&	0.92	&	0.97	&	1.03E-02	&	0.42	&	6.72	&	0.93	&	&	0.40	&	6.84	&	0.97	\\
782	&	3899.4	&	1.51	&	3.33	&	2.70	&	5.07E-06	&	0.61	&	5.55	&	0.0009	&	&	0.34	&	5.06	&	0.002	\\
9032	&	3342.0	&	0.27	&	2.82	&	1.55	&	8.47E-03	&	0.22	&	5.83	&	2.36	&	&	0.21	&	5.87	&	2.50	\\
9069	&	3515.6	&	0.19	&	2.31	&	1.19	&	1.35E-02	&	0.38	&	6.43	&	1.67	&	&	0.31	&	6.52	&	2.01	\\
9077	&	3097.4	&	0.32	&	0.01	&	1.98	&	7.06E-04	&	0.10	&	4.79	&	0.54	&	&	0.14	&	5.13	&	0.40	\\
9108	&	3515.6	&	0.13	&	0.20	&	0.99	&	5.74E-03	&	0.36	&	6.60	&	0.62	&	&	0.32	&	6.77	&	0.69	\\
9171	&	3605.8	&	0.24	&	1.79	&	1.26	&	4.59E-05	&	0.41	&	6.42	&	0.01	&	&	0.38	&	6.41	&	0.01	\\
9175	&	2844.5	&	0.95	&	0.01	&	4.02	&	3.12E-04	&\nodata	&\nodata	&	\nodata	&	&	0.10	&	\nodata	&0.48	\\
9209	&	3801.9	&	0.65	&	3.91	&	1.87	&	3.94E-03	&	0.55	&	6.10	&	0.54	&	&	0.38	&	5.70		&	0.76	\\
9220	&	3097.4	&	0.09	&	0.01	&	1.06	&	5.35E-04	&	0.15	&	6.04	&	0.15	&	&	0.16	&	6.35		&	0.14	\\
9250	&	3845.9	&	0.83	&	2.57	&	2.07	&	-6.98E-04	&	0.57	&	5.93	&	\nodata	&	&	0.37	&	5.53	&	\nodata	\\
9280	&	3258.4	&	0.19	&	1.34	&	1.36	&	1.01E-04	&	0.22	&	6.01	&	0.02	&	&	0.19	&	6.15	&	0.03	\\
9298	&	2930.9	&	0.31	&	0.01	&	2.18	&	3.22E-05	&\nodata	&\nodata	&	\nodata	&	&	0.11	&	4.75	&	0.02	\\
\enddata
\end{deluxetable}

\end{document}